\documentclass[a4paper,fleqn,usenatbib]{mnras}

\usepackage{newtxtext,newtxmath}

\usepackage[T1]{fontenc}
\usepackage{ae,aecompl}

\usepackage[dvipdfmx]{graphicx}	
\usepackage{amsmath}	
\usepackage{amssymb}	
\usepackage{indentfirst}	
\usepackage{color}
\usepackage{pict2e}

\title[Upper mass limit in spherical accretion]{ 
Upper stellar mass limit by radiative feedback at low-metallicities:
metallicity and accretion rate dependence}

\author[Fukushima et al.]{
Hajime Fukushima$^{1,2}$\thanks{E-mail:fukushima@astr.tohoku.ac.jp}
Kazuyuki Omukai$^{1}$
Takashi Hosokawa$^{2}$
\\
$^{1}$Astronomical Institue, Tohoku University, Aoba, Sendai 980-8578, Japan\\
$^{2}$Department of Physics, Kyoto University, Sakyo, Kyoto 6060-8502, Japan
}

\date{Accepted XXX. Received YYY; in original form ZZZ}

\pubyear{2017}

\begin{document}
\label{firstpage}
\pagerange{\pageref{firstpage}--\pageref{lastpage}}
\maketitle

\begin{abstract}
We investigate the upper stellar mass limit set by radiative feedback by the forming star with various accretion rates and metallicities. 
To this end, we numerically solve the structures of both a protostar and its surrounding accretion envelope assuming 
a spherical symmetric and steady flow. 
The optical depth of the dust cocoon, a dusty part of the accretion envelope, differs among the direct light from the stellar photosphere and the diffuse light re-emitted as dust thermal emission.
As a result, varying the metallicity qualitatively changes the way that the radiative feedback suppresses the accretion flow. 
With a fixed accretion rate of $10^{-3}M_{\odot} {\rm yr^{-1}}$, the both direct and diffuse lights  jointly operate to prevent the mass accretion at $Z  \gtrsim 10^{-1}Z_{\odot}$. 
At $Z \lesssim 10^{-1}Z_{\odot}$, the diffuse light is no longer effective, and the direct light solely limits the mass accretion. 
At $Z \lesssim 10^{-3}Z_{\odot}$, the HII region formation plays an important role in terminating the accretion. 
The resultant upper mass limit increases with decreasing metallicity, from a few $\times~10~M_\odot$ to $\sim 10^3~M_\odot$ over 
$Z = 1Z_{\odot} - 10^{-4}~Z_\odot$. 
We also illustrate how the radiation spectrum of massive star-forming cores changes with decreasing metallicity. 
First, the peak wavelength of the spectrum, which is located around $30 \mu {\rm m}$ at $1Z_{\odot}$, shifts to $< 3 \mu {\rm m}$ at $Z \lesssim 0.1 Z_\odot$. 
Second, a characteristic feature at $10\mu\rm{m}$ due to the amorphous silicate band appears as a dip at $1Z_{\odot}$, but changes to a bump at $Z \lesssim 0.1 Z_{\odot}$. 
Using these spectral signatures, we can search massive accreting protostars in nearby low-metallicity environments with up-coming observations. 
\end{abstract}



\section{Introduction}\label{introduction}
Massive stars play pivotal roles in the formation and evolution of galaxies.
They impact mechanical feedback to the interstellar medium, 
such as the formation and development of HII regions, wind-driven bubbles, and supernova remnants, 
which regulates the galactic-scale star formation activities. 
They also control the cosmic chemical evolution
by forging and ejecting heavy elements in supernova explosions.
Their roles would be more significant in young, low-metallicity galaxies.
For instance, copious ionizing photons from massive stars
in such galaxies may have caused the reionization of the universe.
In understanding various aspects of the early structure formation, 
detailed knowledge of massive star formation in low-metallicity environments
is essential.

The standard scenario postulates that the star formation begins with the
cloud collapse which eventually produces an embryo protostar. 
The protostar then accretes the gas from a surrounding envelope, and
grows in mass. The evolution in such a late accretion stage
is vital for understanding the massive star formation \citep[e.g.,][]{ZY07},
because the stellar luminosity also rapidly increases and potentially
impacts the strong feedback which may halt the accretion.
Since the final stellar mass is determined when the accretion ceases, 
understanding of the interplay between the radiative feedback and 
accretion flow is indispensable to derive how massive the star finally becomes.
Regarding the present-day massive star formation, 
previous studies have shown that radiation pressure exerted
on dust grains becomes strong enough to disturb the accretion flow
\citep[e.g.,][]{Larson_Starrfield_1971,Kahn1974,Yorke_Kruge1977}.
In particular, under the assumption of the spherical symmetry,
the accretion is terminated by such an effect before the stellar mass
exceeds $20 M_{\odot}$ with $< 10^{-5} M_{\odot} {\rm {yr^{-1}}}$, i.e.,
typical rates for the low-mass star formation
\citep[so-called ``radiation pressure barrier''; ][]{Wolfire_Cassinelli1987}.
\citet{McKee_Tan2003} proposed that the massive star formation actually occurs
under some non-standard conditions such as massive dense cores supported by turbulence
and/or magnetic fields, where the rapid accretion with
$\dot{M} \gtrsim 10^{-4} M_{\odot} {\rm yr^{-1}}$ is realized.
Recent multi-dimensional simulations also show that
the radiation pressure is less powerful than previously thought, because
photons predominantly escape through less dense parts in the accretion
envelope, e.g., polar directions in the case of the disk accretion
\citep[e.g.,][]{Yorke_Bodenheimer1999,Krumholz2009,Kuiper2010,Klassen2016,Rosen2016}.


In the primordial star formation, on the other hand, there is no 
radiation-pressure barrier because of the lack of dust grains.
A plausible mass limiting mechanism will be due to an HII region created 
around the protostar, as it expands through the accretion envelope 
owing to the excess gas pressure. 
Following pioneering studies assuming the spherical symmetry
\citep[e.g., ][and references therein]{Yorke1986,Omukai_Inutsuka2002} ,
particular attention is paid to how the 
feedback operates in realistic situations such as the disk accretion
in recent studies
\citep[e.g.,][]{Hollenbach1994,McKee_Tan2008,Hosokawa2011,Hosokawa2016}.
\cite{Hosokawa2011} show that, once an HII region begins 
to expand dynamically through the accretion envelope, the gas infalling toward the disk 
is blown away, so that the mass supply onto the protostar is eventually suppressed. 
Note that the expansion of the HII region is driven by its {\it gas pressure}
excess, and thus the resulting feedback essentially differs from that
caused by the {\it radiation pressure} described above. 
The UV feedback potentially works also in the present-day cases
 \citep[e.g.,][]{YW96}, but it is still uncertain how the radiation-pressure 
 barrier and UV feedback jointly operate.


Although still in debate, the studies above clearly suggest that
the radiative feedback works differently among the present-day and primordial 
massive star formation, implying its variations over metallicities.
\citet{Hosokawa_Omukai2009b} (hereafter HO09b) presented a study on
such metallicity-dependence with a full coverage of $0 \leq Z \leq Z_\odot$
assuming the spherical symmetry.
They estimate the upper mass limits set by each feedback effect
with simple analytic arguments.
The accretion histories onto the protostar are obtained from the thermal
evolutionary tracks of a collapsing core at various metallicities \citep{Omukai2005}.
Their results are briefly summarized as follows.
Since the average accretion rate is higher with the lower metallicity, 
the resulting protostellar evolution also differs with different metallicities.
In addition, the accretion envelope has lower 
dust opacity for lower metallicity, the dominant 
feedback mechanism changes depending on metallicity.
H09b have derived the upper mass limit as a function of the
metallicity, which are set by the radiation pressure on dust grains
for $Z \gtrsim 10^{-3}~Z_{\odot}$ and by the expansion
of the HII region for $Z \lesssim 10^{-3}~Z_{\odot}$.


In this paper, we further improve our treatments to derive
the metallicity-dependent upper stellar mass limits set by
the protostellar feedback.
First, our calculation covers a much broader range of the different accretion rates
for a given metallicity, reflecting the fact that such diversity is
commonly expected at both the solar and zero metallicities
\citep[e.g.,][]{McKee_Tan2003,Hirano15}.
Second, in addition to the analytic approach employed in HO09b, we also investigate
the radial structure of the accretion envelope by numerically solving
the frequency-dependent radiation transport consistently through it.
This allows us to derive radiation spectrum of massive star-forming
cores at low metallicities, with which we predict their observational signatures.
For instance, a low-metallicity massive protostar exhibits a 
broader and harder radiation spectrum than its solar-metallicity counterpart.  
Such characteristic features will be useful for future
observational surveys for nearby low-metallicity
massive protostars.


We organize the rest of the paper as follows. 
In Section \ref{stellar.evlol}, we first overview the protostellar evolution 
under different accretion rates and metallicities. 
We then analytically estimate the upper mass limits set by radiative feedback 
for a wide range of the accretion rates and metallicities in Section \ref{sec.kaiseki}. 
Next, we construct numerical models of an accretion envelope around the protostar,
through which the frequency-dependent radiative transfer is solved. 
Sections \ref{sec.method_env} and \ref{sec.kekka_env} are devoted 
to describe the method and results of such numerical modeling. 
In Section  \ref{sec.tenkei}, we apply our numerical modeling to
the typical accretion histories expected at each metallicity, 
and compare the results to our previous work H09b. 
Finally, our basic results are summarized with related discussions in
Section \ref{sec.diss}. We also briefly describe 
methods used in our numerical modeling in Appendixes \ref{apdi1} and \ref{apdi3}.


\section{Protostellar evolution}\label{stellar.evlol}

\begin{figure}
	\begin{center}
		\includegraphics[width=70mm]{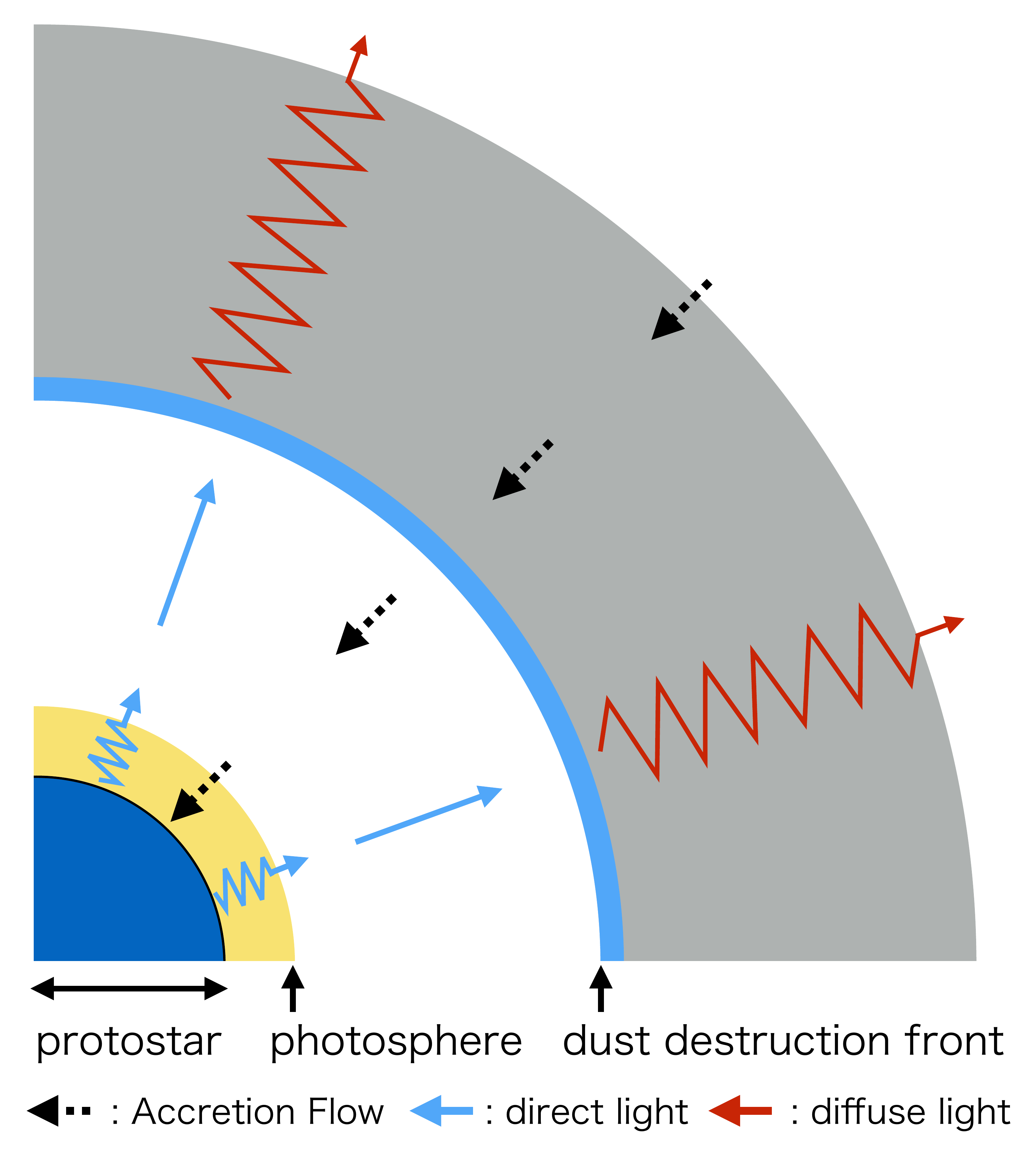}
	\end{center}
\caption{A schematic view of an accretion envelope around a protostar with 
non-zero metallicity and high accretion rate, i.e., 
$Z=Z_\odot$ and $\dot M_* = 10^{-4}~M_{\odot} {\rm yr^{-1}}$. 
Since the innermost dust-free accretion flow becomes optically thick before hitting
the stellar surface, the photosphere appears within the accretion flow.
Dust grains are only contained in the outer part beyond the 
dust destruction front, where the grain temperature reaches the
sublimation value.
The UV and optical light directly coming from the photosphere is 
absorbed in a thin layer at the dust destruction front, 
and re-emitted as the diffuse IR light.
}
\label{fig0213.1}
\end{figure}

\begin{figure}
	\begin{center}
		\includegraphics[width=\columnwidth]{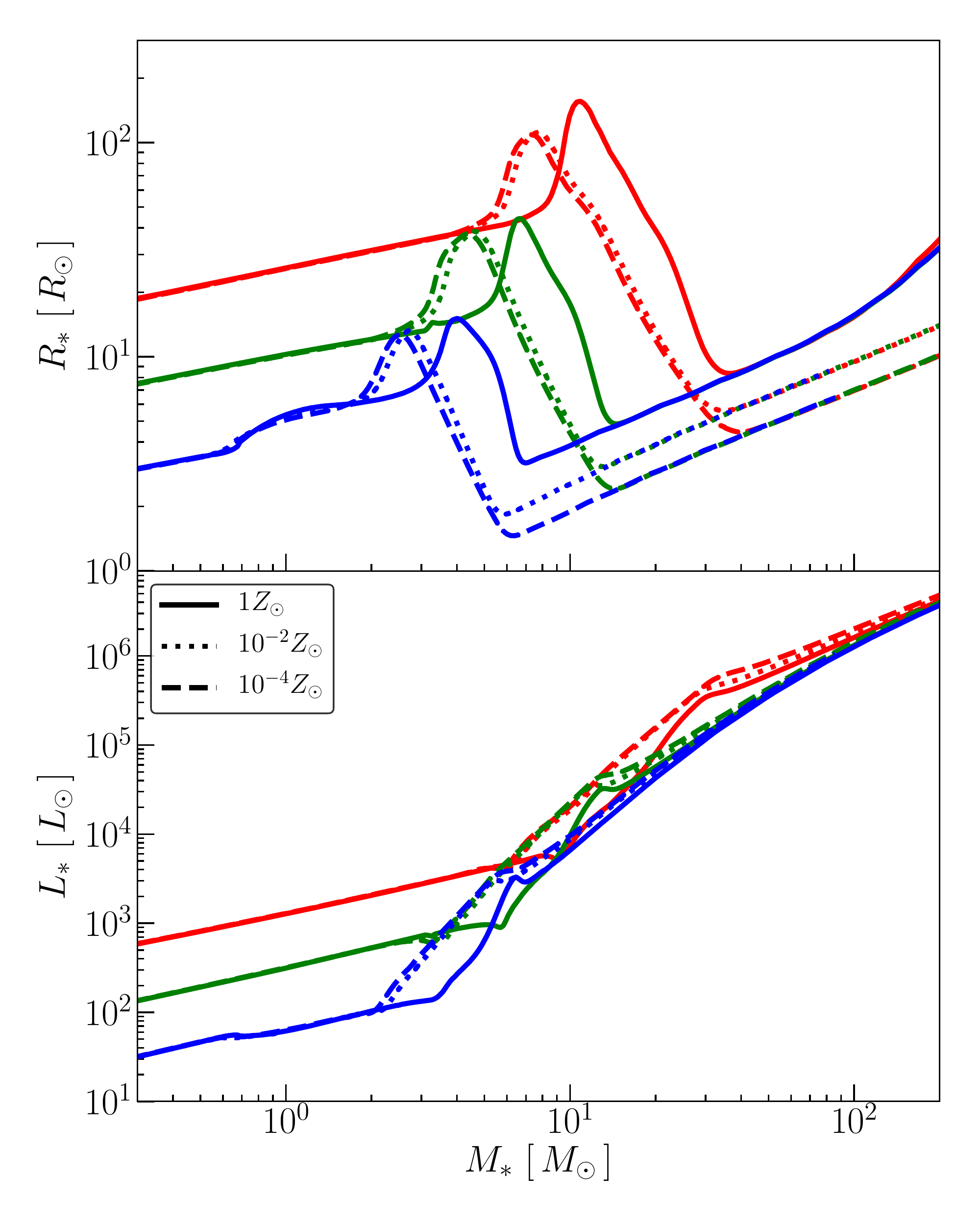}
	\end{center}
\caption{The protostellar evolution with different 
metallicities and accretion rates. 
The top and bottom panels show the evolution of the stellar radius and luminosity respectively 
against the stellar mass. 
In each panel, the red, green, and blue curves represent the 
different accretion rates of $\dot M = 10^{-3}$, $10^{-4}$, and $10^{-5}~M_{\odot} {\rm yr^{-1}}$.
For each accretion rate, the solid, dotted, and dashed lines
are for the metallicities of $Z = 1$, $10^{-2}$, and $10^{-4}Z_{\odot}$. 
}
	\label{fig0119.1}
\end{figure}

The strength of the protostellar radiation feedback depends on
the stellar luminosity and effective temperature, which are given by
solving the stellar interior structure. We have developed
numerical models of the protostellar evolution in our previous studies
\citep[e.g.,][for full details]{Hosokawa_Omukai2009a}, which are briefly
reviewed in what follows.


 Figure \ref{fig0213.1} schematically shows the structure of an
accretion envelope around a protostar at non-zero metallicity. 
The outermost part is the dust cocoon, where the gas falls toward the
star together with dust grains. 
We assume that, throughout the paper, dust 
grains are perfectly coupled with the gas in their motion. 
In an inner side of the dust cocoon, the grains are heated
up by the stellar irradiation and sublimate at the
dust destruction front. 
The envelope is dust-free within this radius, 
where the gas is mostly optically thin for the stellar light 
owing to the small gas opacity. In a very vicinity of the star, 
however, even the dust-free flow becomes optically thick if the accretion rate is
high enough. The photosphere appears where the gas is still in
free-fall. For such a case, the position of the photosphere 
differs from that of the stellar surface, where the accreting gas
hits on the star forming a shock front.


In the calculations of the protostellar evolution, we only solve the structure
of the stellar interior and, if any, optically-thick part of the dust-free accretion flow.
The structure of the outer accretion envelope is separately considered
from the inner part (Section \ref{sec.method_env}).
 We solve the so-called four stellar structure equations, i.e., continuity, hydrostatic
balance, energy conservation and transport with taking account of effects of the
mass accretion \citep[e.g.,][]{Hosokawa_Omukai2009a}.
The outer boundary conditions are provided by the jump condition at the
surface shock front \citep{Stahler1980_2,Hosokawa_Omukai2009a}.
If the dust-free flow remains optically thin until hitting the surface, we just
assume a free-fall profile. Otherwise, we guess the position of the photospheric radius
 $R_{\rm ph}$ that satisfies $\rho \kappa R_{\rm ph}=1$, and solve the flow structure
for $r < R_{\rm ph}$ using the diffusion approximation for the radiation transport.
We construct a model which satisfies the outer boundary condition
by an iterative shooting method \citep[e.g.,][]{Stahler1980_2,Hosokawa_Omukai2009a}.
Assuming the free-fall for the optically-thin part
is not accurate because in the dust cocoon the flow may be
substantially decelerated by the radiation pressure to deviate from the free-fall flow
(see Section \ref{sec.kekka_env} below). 
Even for such a case, however, the gas is accelerated again
by the gravitational pull once passing through the dust destruction front 
because the opacity is substantially reduced. 
As a result, the free-fall velocity is easily recovered. 


We start an evolutionary calculation with an initial model with $M_{*} = 0.05M_{\odot}$
constructed following \citet{Stahler1980}.
With a high accretion rate such as $10^{-3} M_{\odot} {\rm yr^{-1}}$, 
we adopt the larger initial
mass $0.1 M_{\odot}$ to facilitate the numerical convergence of the iterative process.
The protostellar evolution does not differ with different initial models
except in an early stage immediately after starting calculations.


 The evolution of the stellar luminosity and radius are shown
in Figure \ref{fig0119.1} for some representative cases.
Note that the total luminosity $L_{*}$ is the sum of the 
stellar interior luminosity  $L_{\rm int}$ and accretion luminosity $L_{\rm acc}$.
The protostellar evolution is divided into four characteristic stages
\citep[e.g.,][]{Hosokawa_Omukai2009a}:
(1) the adiabatic accretion ($M_{*} \lesssim 4M_{\odot}$),
(2) swelling ($4M_{\odot} \lesssim M_{*} \lesssim 6M_{\odot}$),
(3) Kelvin-Helmholtz contraction ($6M_{\odot} \lesssim M_{*} \lesssim 15M_{\odot}$),
and (4) main sequence accretion ($M_{*} \gtrsim 15M_{\odot}$),
where the mass ranges are for the case with $\dot M = 10^{-4} M_{\odot} {\rm yr^{-1}}$ and $Z = 1Z_{\odot}$.
With lower metallicity, the transitions among the stages occur at lower stellar masses,
except the arrival to the stage (4).
The evolution in each stage is briefly described as follows:


\begin{itemize}
 \item[(1)] {\it adiabatic accretion}:
In the early stage where the protostar mass is small, the accretion time
$t_{\rm acc} = M_{*}/\dot M$ is smaller than Kelvin-Helmholtz time
$t_{\rm KH} = (GM_{*}^2)/(RL_{\rm int})$.
The protostar evolves adiabatically because the amount of entropy brought
into the star with the accretion flow is larger than that emitted by radiation.
The evolution of the radius $R_{*}$ and luminosity $L_{*}$ shows
no metallicity-dependencies.


 \item[(2)] {\it swelling}: The stellar radius temporarily increases
to exceed $100 R_{\odot}$ at maximum,  which is caused by the outward
transport of the accumulated entropy in the stellar deep interior by the so-called luminosity wave \citep{SPS86}.
With the same accretion rate, the swelling occurs at lower stellar mass
with lower metallicity (top panel of Figure \ref{fig0119.1}).


 \item[(3)] {\it Kelvin-Helmholtz contraction}: With the timescale balance of
$ t_{\rm acc} \geq t_{\rm KH}$, the protostar is able to emit the internal energy
by radiation and then contracts.
This contraction continues until the interior temperature 
becomes high enough to commence hydrogen burning.


 \item[(4)] {\it main sequence accretion}:
After the star enters this stage where the hydrogen burning takes place
near the stellar center, the protostar radius becomes independent of the accretion rates.
The low-metallicity stars cannot achieve efficient hydrogen burning via the CNO cycle
with its original small abundance of metals.
The protostar then continues to contract until CNO elements are additionally
provided by the 3-$\alpha$ reactions.
At the lower metallicities, the radius of the main sequence star is thus smaller
than its solar-metallicity counterpart.
\end{itemize}

\section{Analytical estimate for upper mass limits}\label{sec.kaiseki}

 We first analytically estimate the upper stellar mass limits
for various metallicities $Z$ and accretion rates $\dot M$, using the protostellar
evolution presented in the previous section.
We consider the two feedback effects caused by
the radiation pressure exerted on dust grains,
and the formation of an HII region by ionizing photons.


Our calculation covers a broad range of the metallicity of $0 \leq Z \leq 1~Z_\odot$.
The mean accretion rate onto the protostar scales approximately
with the temperature of a collapsing dense core as
	\begin{eqnarray}
			\dot M \sim \frac{M_{\rm J}}{t_{\rm ff}}
\sim  \frac{ c_{\rm s}^3}{G}
= 2.5 \times 10^{-6} \left( \frac{T}{10 \rm{K}} \right)^{3/2} M_{\odot} {\rm yr^{-1}},
\label{1117.1}
	\end{eqnarray}
 where $M_{\rm J}$ , $t_{\rm ff} = \sqrt{{3 \pi}/{32 G \rho}}$ , $c_{\rm s}$
and $T$ are the Jeans mass, the free fall time, the sound speed and the gas temperature.
Since the typical core temperature varies over the range
$10~{\rm K} \lesssim T \lesssim 10^{3}\rm{K}$ depending on metallicity,
the corresponding accretion rate covers 
$10^{-5} M_{\odot} {\rm yr^{-1}} \lesssim \dot M \lesssim  10^{-3} M_{\odot} {\rm yr^{-1}}$
\citep[][HO09b]{Omukai2005}.
In this paper, we also consider that, even at the same metallicity, 
the mean accretion rates can be different to  some extent.
Recall that very rapid accretion with $\gtrsim 10^{-4}~M_\odot {\rm yr^{-1}}$
is postulated in high-mass star formation in the Milky Way, 
where the typical value for low-mass star formation 
is less than $10^{-5}~M_\odot {\rm yr^{-1}}$ \citep[e.g.,][]{McKee_Tan2003}.
In order to take such additional diversity into account,
we examine the cases with different accretion rates 
$\dot M = 10^{-3}$, $10^{-4}$, and $10^{-5} M_{\odot} {\rm yr^{-1}}$ 
for each metallicity.

\subsection{Optical depth of dust cocoon}\label{sec.optical_depth}

In order to investigate the feedback caused by the radiation pressure,
we first estimate the optical depth of the dust cocoon at a frequency $\nu$ 
by integrating outward from the dust destruction radius $R_{\rm d}$ as
\begin{eqnarray}
			\tau_{\nu} &=& \int ^{\infty}_{R_{\rm d}} \kappa_{\rm \nu} \rho d r = \frac{\dot M \kappa_{\rm \nu}}{2 \pi \sqrt{2 G M_{*}}} R_{\rm d} ^{-1/2}, \label{0123.3}
\end{eqnarray}
where we have used the continuity equation for spherically symmetric steady flow
	\begin{eqnarray}
			\rho = \frac{\dot M}{4 \pi r^2 |u|}, \label{eq1.6.1}
	\end{eqnarray}
and assumed the free-fall velocity for $u$ as
	\begin{eqnarray}
		u =  u_{\rm ff} = \sqrt{\frac{2 G M_{*}}{r}} . \label{0331.1}
	\end{eqnarray}

The radial position of the dust destruction front $R_{\rm d}$, which is still
unknown in Equation \eqref{0123.3}, is estimated as follows.
We assume that the temperature of dust grains is determined by the balance between
the absorption and emission of photons.
Since the dust evaporates at the sublimation temperature $\simeq 1200 {\rm K}$,
such energy balance at $r = R_{\rm d}$ is given by
	\begin{eqnarray}
		\int \kappa_{\nu} J_{\nu}(R_{\rm d} ) d \nu = \int \kappa_{\nu} B_{\nu} (1200 \rm{K}) d \nu. \label{0123.6}
	\end{eqnarray}
Neglecting small opacity between the stellar photosphere and the dust destruction front,
the mean intensity $J_{\nu}(r)$ is written as
\begin{eqnarray}
			J_{\nu}(r) = \frac{B_{\nu} \left(T_{\rm eff} \right)}{2} \left( 1 - \sqrt{1 - \left(\frac{R_{\rm{ph}}}{r} \right)^2} \right) ,\label{0123.5}
\end{eqnarray}
where $R_{\rm{ph}}$ is the stellar photosphere radius, and $T_{\rm eff}$ is effective temperature.
We can derive $R_{\rm d}$ by substituting Equation \eqref{0123.6} into \eqref{0123.5},
	\begin{eqnarray}
		R_{\rm d} = \frac{R_{\rm{ph}}}{2 \sqrt{x_{\rm d}(1 - x_{\rm d})}} \simeq 110 {\rm{AU}} \left( \frac{L_{*}}{10^{5}L_{\odot}} \right)^{1/2} \left( \frac{\kappa_{\rm P} \left( T_{\rm eff} \right)}{400 \rm{cm^{2}g^{-1}}} \right)^{1/2}, \label{0123.7}
	\end{eqnarray}
where the Planck mean opacity $\kappa_{\rm P}$ is defined by
	\begin{eqnarray}
		\kappa_{\rm P} (T) = \frac{\int ^{\infty}_{0} \kappa_{\nu} B_{\nu} (T)  d \nu}{\int ^{\infty}_{0} B_{\nu}(T) d\nu} ,\label{0330.2}
	\end{eqnarray}
and $x_{\rm d}$ is
	\begin{eqnarray}
		x_{\rm d} = \left( \frac{1200 \rm{K}}{T_{\rm eff}} \right)^4  \frac{\kappa_{\rm P}(1200\rm{K})}{\kappa_{\rm P}(T_{\rm eff})}, \label{0123.8}
	\end{eqnarray}
which is well approximated as $x_{\rm d } \ll 1$ because the effective temperature
$T_{\rm eff}$ is sufficiently higher than $1200 \rm{K}$.

\subsection{Suppression mechanisms of mass accretion}\label{analytical_upper_mass}
\begin{figure}
	\begin{center}
		\includegraphics[width=80mm]{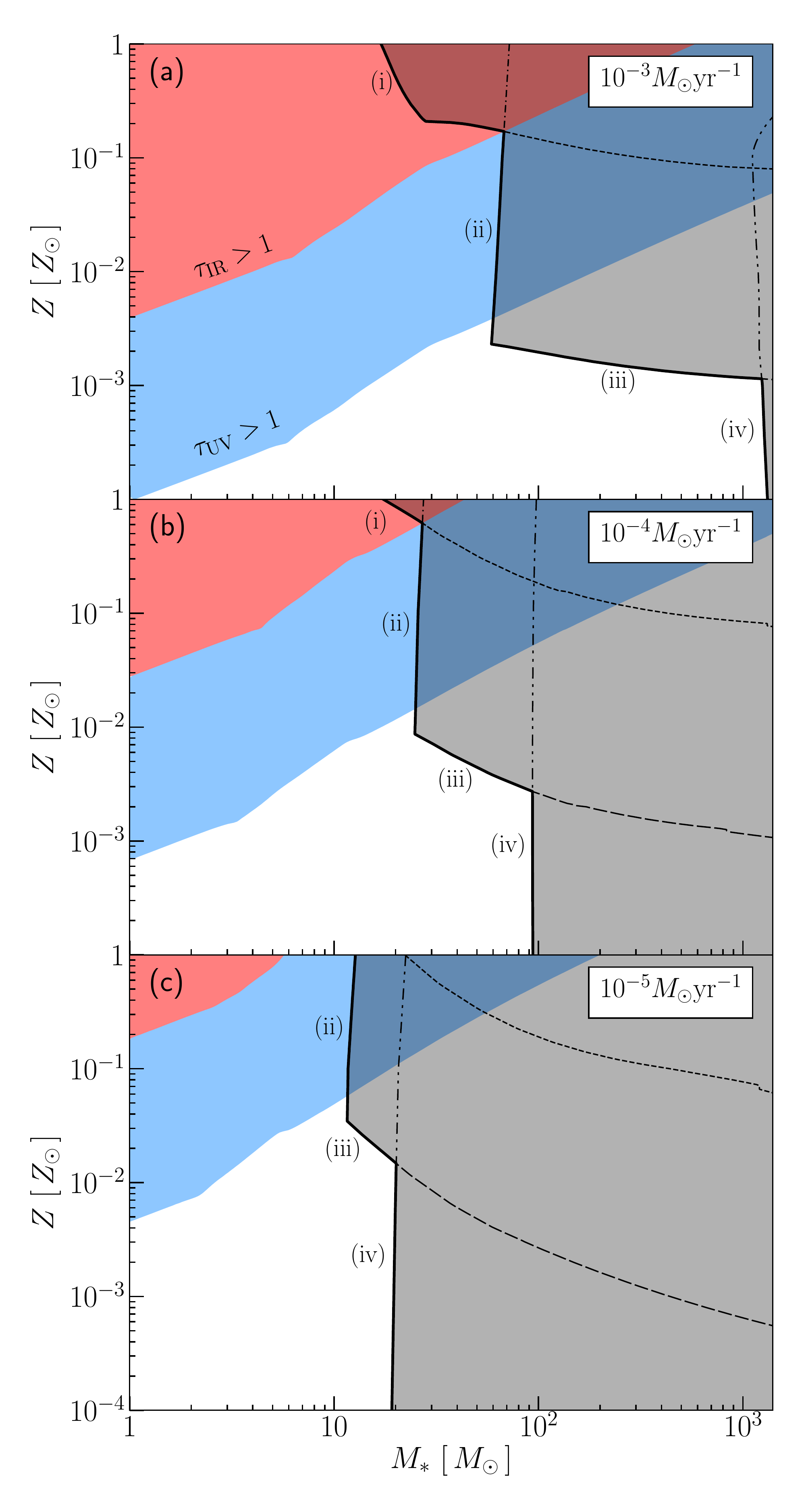}
	\end{center}
\caption{The analytically-estimated upper mass limits
with different constant accretion rates $\dot M = 10^{-3}~M_{\odot} {\rm yr^{-1}}$ (top panel),
$10^{-4}~M_{\odot} {\rm yr^{-1}}$ (middle panel), and $10^{-5}~M_{\odot} {\rm yr^{-1}}$ (bottom panel).
In each panel, the solid line shows the variation of the upper
mass limits with different metallicities: 
(i) the radiation pressure by the diffuse light terminates the mass accretion; 
(ii) the flow is decelerated mostly by the direct radiation pressure around the dust destruction front;
(iii) the radiation pressure by the direct light onto the entire cocoon decelerates the flow;
(iv) the stellar mass is limited by the growth of the HII region.
The contributions by individual feedback effects are also
plotted with the different lines in the grey-shaded area.
The red- and blue-shaded areas denote that the optical
depth of the dust cocoon exceeds the unity for the
diffuse and direct lights. 
}
	\label{fig0124.1}
\end{figure}

\subsubsection{Radiation pressure in dust cocoon}
\label{dust_press}

 If the dust cocoon is optically thick for the stellar irradiation, most of
photons coming from the stellar photosphere are absorbed and re-emitted
as the dust thermal emission. 
Hereafter we call the former as {\it{the direct
light}}, and the latter as {\it{the diffuse light}}.
The optical depths of the dust cocoon for such direct and diffuse lights
control how the radiation pressure affects the dusty accretion flow.
For the direct light, we use the opacity at the frequency
$\nu_{\rm max} = 1.18 \times 10^{15} \rm{Hz}$
(or the wavelength $\lambda _{\rm max} = 0.254 {\rm \mu m}$),
where the Planck function with the typical stellar effective temperature,
$2 \times 10^4 {\rm{K}}$, takes the maximum value.
 For the diffuse component, on the other hand,
the Rosseland mean opacity at $T = 500$~K provides a good approximation.
The numerical values for the direct and diffuse lights are
$\kappa_{\rm UV} = 350 (Z/Z_{\odot}) {\rm {cm^2 g^{-1}}}$ and
$\kappa_{\rm IR} = 5 (Z/ Z_{\odot}) {\rm {cm^2 g^{-1}}}$ respectively,
i.e., the former is 70 times larger than the latter.


 The conditions for suppressing the mass accretion by the radiation pressure
can be classified as in the following (i), (ii), and (iii), depending on whether
the optical depths of the dust cocoon for the direct and diffuse lights,
$\tau_{\rm UV}$ and $\tau_{\rm IR}$, exceed the unity or not:
\begin{itemize}


 \item[(i)] $\tau_{\rm UV},\tau_{\rm IR} > 1$: \\
In this case, the direct light is absorbed and re-emitted in a thin layer
near the dust destruction front.
Since the dust cocoon is also optically thick for the diffuse light,
 the re-emitted IR photons proceed outward via diffusion.
The total momentum received by the dust cocoon is
\begin{eqnarray}
		\int ^{\infty}_{R_{\rm d}} \frac{ \kappa_{\rm IR} L_{*} }{4 \pi r^2 c} \left( 4 \pi r^2 \rho \right) dr = \frac{L_{*}}{c} \int^{\infty}_{R_{\rm d}} \kappa_{\rm IR} \rho d r  = \frac{L_{*} \tau_{\rm IR}}{c},  \label{0331.3}
\end{eqnarray}
while the momentum carried by the absorbed direct light is $L_{*}/c$.
Hence, with $\tau_{\rm IR}>1$, the radiation pressure
caused by the diffuse light dominates the contribution from the direct light.
The mass accretion is terminated when the stellar luminosity exceeds
the Eddington value defined with the IR opacity $\kappa_{\rm IR}$ (HO09b),

\begin{eqnarray}
  L_{*} > L_{\rm Edd,IR} = \frac{4 \pi c G M_{*}}{\kappa_{\rm IR}}. \label{0123.1}
\end{eqnarray}


 \item [(ii)] $\tau_{\rm UV} > 1, \tau_{\rm IR} < 1$: \\
Although the direct light is absorbed near the dust
destruction front as in case (a),
the re-emitted IR diffuse light now escapes through the cocoon freely.
The accretion flow receives the radiation
pressure dominantly from the direct light.
In a spherical shell near the dust destruction front where the direct light
is totally absorbed, the accretion flow receives the momentum flux
$L_{*}/(4 \pi R_{\rm d}^2 c)$ from the radiation.
The accretion flow will be in stall if this momentum flux exceeds that
by the accretion flow $\rho u^2$ \citep{Wolfire_Cassinelli1987},
which is written as
\begin{eqnarray}
	\frac{L_{*}} {4 \pi R_{\rm d}^2 c} > \rho u ^2. \label{0122.1}
\end{eqnarray}
Substituting equations \eqref{eq1.6.1} and \eqref{0331.1},
equation \eqref{0122.1} becomes
\begin{eqnarray}
 \dot M < \frac{L_{*}}{u c}
= \frac{L_{*}}{c} \left( \frac{2 G M_{*}}{R_{\rm d}} \right)^{1/2} .
\label{0122.2}
\end{eqnarray}


\item[(iii)] $\tau_{\rm UV}, \tau_{\rm IR} < 1$: \\
Because the dust cocoon is optically thin, the direct light travels through
the dust cocoon without being absorbed.
In this case, the mass accretion can be hindered if the stellar luminosity
$L_{*}$ exceeds the Eddington value defined with the UV opacity
$\kappa_{\rm UV}$ (HO09b), which is written as
\begin{eqnarray}
 L_{*} > L_{\rm Edd,UV} =  \frac{4 \pi c G M_{*}}{\kappa_{\rm UV}}. \label{0123.2}
\end{eqnarray}

\end{itemize}

\subsubsection{HII region formation}\label{HII_region}

 Ionizing photons emitted from a protostar create an HII region in its surroundings.
When the HII region expands dynamically, mass accretion is terminated 
by the pressure gradient. The epoch when such strong feedback is impacted 
can be estimated by measuring the size of the HII region as follows.


First, the stellar emissivity of ionizing photons $S (\rm{sec^{-1}})$ is given by
\begin{eqnarray}
  S = 4 \pi R_{*}^2 \int ^{\infty} _{\nu_{\rm{L}}}
\frac{\pi B_{\nu}(T_{\rm{eff}})}{h \nu} d \nu ,
\label{eq4.1}
\end{eqnarray}
where $T_{\rm eff}$, $R_{*}$, and $\nu_{\rm{L}}$ are the stellar
effective temperature, radius, and the Lyman limit frequency.
The radius of the HII region $R_{\rm HII}$ is determined by
setting the supply rate of the ionizing photons
equal to the consumption rate \citep[e.g.,][]{Omukai_Inutsuka2002},
\begin{eqnarray}
  S = \int^{R_{HII}}_{R_{*}} \alpha _{\rm B}
n ( {\rm{H}^{+}} ) n ( {\rm{e}^{-}} ) d V + \frac{\dot M}{\mu m_{\rm H}} ,
\label{eq4.2}
\end{eqnarray}
where $\alpha_{\rm B}$ is the case-B recombination coefficient.
In Equation \eqref{eq4.2}, the first term on the right hand side represents
the recombination rate of ionized hydrogen, and the second term is the supply
rate of neutral hydrogen carried into the HII region by the accretion flow.
Since the second term is much smaller than the first term
with the accretion rates considered in this paper, we ignore this term.
The radial density distribution within the HII region,
which is necessary in performing the integration in equation \eqref{eq4.2},
is calculated in a following way.
Since the gas receives the radiation pressure via the electron
scattering within the HII region, the equation of motion is given by
\begin{eqnarray}
 u\frac{du}{dr} = - \frac{GM_{\ast}}{r^2} \left( 1 - \Gamma \right) ,
 \label{eq4.3}
\end{eqnarray}
where $\Gamma =  L _{*}/ L_{\rm{Edd}}$,
and $L_{\rm{Edd}} = 4 \pi c G M_{\ast} / \kappa_{\rm{sc}}$ is the Eddington luminosity
defined with the electron scattering opacity.
We derive the density profile within the HII region using Equations
\eqref{eq1.6.1}, \eqref{0331.1}, and \eqref{eq4.3},
\begin{eqnarray}
 \rho = \frac{\dot M}{4 \pi r^2}
\left[   \frac{2 G M_{\ast} }{R_{\rm{HII}}} \Gamma
       + \frac{2 G M_{\ast}}{r} \bigl( 1 -  \Gamma \bigr)
\right], \label{eq4.5}
\end{eqnarray}
where we have assumed that the accretion flow is well described
as the free-fall outside the HII region.
Substituting Equation \eqref{eq4.5} into Equation \eqref{eq4.2} using
$n({\rm{H}^{+}}) = n({\rm{e}^{-}}) = \rho / \mu  m_{\rm H}$ for the fully ionized gas,
we obtain
\begin{eqnarray}
			S = \frac{\alpha_{\rm B} \dot M^2} {8 \pi \mu^2 m_{\rm {H}}^2 G \left(1 -  \Gamma \right) M_{\ast}} 
\ln \left( \Gamma + \frac{R_{\rm HII}}{R_{*}} (1 - \Gamma) \right) , \label{eq4.7}
\end{eqnarray}
which is further transformed as
\begin{eqnarray}
			R_{\rm HII} = R_{*} \left\{ \exp \left[ \frac{8 \pi \mu^2 m_{\rm H}^2 G M_{\ast}}
{\alpha_{\rm B} \dot M ^2} \left(1 - \Gamma \right) S \right] - \Gamma \right\} \bigg{/} \left( 1 - \Gamma \right).  \label{eq4.8}
\end{eqnarray}
This equation suggests that the HII region rapidly expands
once the stellar emissivity of ionizing photons exceeds a critical value \citep{Yorke1986,Omukai_Inutsuka2002},
\begin{eqnarray}
  S \simeq
    \frac{\alpha_{\rm B} \dot M^2}
         {8 \pi \mu^2 m_{\rm H}^2 G M_{\ast} \left( 1 - \Gamma \right)}.
\label{eq4.10}
\end{eqnarray}


As a critical size of the HII region, we consider the gravitational
radius for the ionized gas
\begin{eqnarray}
r_{\rm g} \equiv \frac{GM_{*}}{c_{\rm HII}^2},
\end{eqnarray}
where $c_{\rm HII}$ is the sound velocity of the photoionized gas.
When the HII region is smaller than
the gravitational radius, the gas plunges into the HII
region at the supersonic velocity.
The presence of the large pressure excess of the HII
region never propagates to the outer neutral flow,
so that the accretion is not hindered.
Once the radius of the HII region exceeds the
gravitational radius, however, the accretion flow
has only subsonic velocity at the ionizing front.
At this moment, the HII region begins to expand dynamically
forming a preceding shock front, blowing away
the accreting gas.
Therefore, the condition for terminating the accretion
by the formation of the HII region is given by
\begin{eqnarray}
  R_{\rm HII } > r_{\rm g}.
\label{0330.2.1}
\end{eqnarray}


 In this paper, in order to calculate $R_{\rm HII}$ and $r_{\rm g}$, we assume that the HII region temperature is $10^{4} \rm{K}$ , and use the recombination rate of hydrogen $\alpha_{\rm B} = 2.6 \times 10^{-13} \rm{cm^3 s^{-1}}$ \citep{Hummer1987}, the electron scattering opacity $\kappa _{\rm sc} = 0.34\rm{cm^{2}g^{-1}}$ and $\mu m_{\rm H} = 2.2 \times 10^{-24} \rm{g}$.

\subsection{$Z$- and $\dot M$-dependence of upper stellar mass limits}

 Figure \ref{fig0124.1} shows
the analytically-estimated upper stellar mass limits for each metallicity
with different accretion rates $\dot M =10^{-3}$, $10^{-4}$, and
$10^{-5} M_{\odot}\rm{yr^{-1}}$.
In each panel, the solid line represents the upper limits set by
the feedback, i.e., the radiation pressure exerted on dust grains
or the formation of an HII region. The grey shaded area indicates
that the steady accretion is not allowed by such feedback effects.
We can also see the contributions from the individual feedback effects:
(i) radiation pressure on to the entire cocoon by the diffuse light
(Eq.~\ref{0123.1}, short dashed lines),
(ii) radiation pressure on to the dust destruction front
(Eq.~\ref{0122.2}, dot-dashed lines),
(iii) radiation pressure on to the entire cocoon by the direct light
(Eq.~\ref{0123.2}, long dashed lines),
and the formation of the HII region (Eq.~\ref{0330.2.1},
dot-dot-dashed lines).


First, we focus on the cases with the accretion rate
$\dot M = 10^{-3} M_{\odot} {\rm yr^{-1}}$ (Fig.~\ref{fig0124.1}-a),
which occur in the present-day high-mass star formation or
low-metallicity environments where the accretion envelope
has relatively high temperature.
In this case, for $Z \gtrsim 0.1~Z_{\odot}$, the radiation pressure caused
by the diffuse light (e.g., Eq.~\ref{0123.1}) is most effective
to limit the stellar mass.
The upper mass limit is $\simeq 20~M_{\odot}$ at $Z = 1~Z_{\odot}$,
which increases with decreasing metallicity
because the cocoon's IR optical depth decreases.
At the lower metallicities of
$4 \times 10^{-3}~Z_{\odot} \lesssim Z \lesssim 0.1~Z_{\odot}$,
the radiation pressure at the dust destruction front
provides the most stringent upper limits, which do not
vary with $Z$ at $M_* \simeq 60 M_{\odot}$.
The dust cocoon becomes optically thin for the stellar direct light
even for $M_* \lesssim 60 M_{\odot}$ at even lower metallicities
in the range
$10^{-3}~Z_{\odot} \lesssim Z \lesssim 4 \times 10^{-3}~Z_{\odot}$.
The direct light is not absorbed at the dust destruction front
any more, but travels freely through the dust cocoon.
The radiation pressure by the direct light over the whole cocoon
causes the strongest feedback at such metallicities.
At $Z < 10^{-3}Z_{\odot}$, where the radiation pressure on dust grains
becomes ineffective, the HII region formation is the
primary feedback mechanism to terminate the mass accretion.
In this case, the upper mass limit is about $10^{3}M_{\odot}$
from Equation \eqref{0330.2.1}, and is hardly dependent on the metallicity.
In summary, with the accretion rate $\dot M = 10^{-3} M_{\odot} {\rm yr^{-1}}$,
we can divide the metallicity range into three regions according to the dominant feedback mechanism
: $0.1~Z_{\odot} \lesssim Z$ with the
radiation pressure by the diffuse light through the dust cocoon,
$10^{-3}~Z_{\odot} \lesssim Z \lesssim 0.1~Z_{\odot}$ with the
radiation pressure by the direct light at the dust destruction front,
and $Z \lesssim 10^{-3}Z_{\odot}$ with the HII region formation.
We have ignored the narrow range where the radiation pressure
through the cocoon by the direct light is most effective.
The upper mass limit changes almost stepwise as
$M_* \simeq 20M_{\odot}$, $60M_{\odot}$, and $10^{3}M_{\odot}$
over these three ranges of metallicity.


 Next, we consider the cases with the ten times lower accretion rate
$\dot M = 10^{-4} M_{\odot} {\rm yr^{-1}}$ presented in Figure \ref{fig0124.1} (b).
As in panel (a), the upper mass limit varies almost
stepwise over different metallicities:
$\simeq 20M_{\odot}$ ($Z \gtrsim 0.6 Z_{\odot}$),
$30M_{\odot}$ ($10^{-2}Z_{\odot} \lesssim Z \lesssim 0.6 Z_{\odot}$),
and $90M_{\odot}$ ($Z \lesssim 10^{-2}Z_{\odot}$).
Mechanism of terminating the mass accretion changes with decreasing
metallicity, in the same ordering as in the case of $\dot M = 10^{-3} M_{\odot} {\rm yr^{-1}}$.
At a fixed metallicity and stellar mass, the optical depths over the
dust cocoon $\tau_{\rm UV}$ and $\tau_{\rm IR}$ are lower for
the lower accretion rate. 
Therefore, the dominant feedback mechanism
switches each other at higher metallicity than in the case of $\dot M = 10^{-3} M_{\odot} {\rm yr^{-1}}$
shown in panel (a).
We also see that the upper mass limit is somewhat
smaller than in panel (a), especially for $Z \lesssim 0.6~Z_\odot$,
reflecting the $\dot M$-dependencies in the conditions
given by equations \eqref{0122.2} and \eqref{0330.2.1}.


  In the same manner, Figure \ref{fig0124.1} (c) shows the cases
with $\dot M = 10^{-5} M_{\odot} {\rm yr^{-1}}$.
In this case, at even $1Z_{\odot}$,
the dust cocoon is optically thin for the diffuse light
(i.e., $\tau_{\rm IR} < 1$) 
when the radiation feedback becomes strong enough to limit the accretion. 
As a result, the full range of the metallicity is divided into the two
regions where the upper mass limits are $\simeq 10M_{\odot}$
for $Z \gtrsim 5 \times 10^{-2}Z_{\odot}$ and $20M_{\odot}$
for $Z \lesssim 5 \times 10^{-2}Z_{\odot}$.
With such a accretion rate, the formation of massive stars with
$M_* \gtrsim 20 M_{\odot}$ is not allowed by the feedback effects
even at the lowest metallicity $Z \lesssim 10^{-4}~Z_\odot$.


 Note that there is a critical metallicity $Z_{\rm limit}$, below which
the radiation pressure becomes ineffective to terminate the mass accretion.
Since the stellar luminosity never exceeds the usual Eddington
value defined with the electron scattering opacity,
radiation pressure on dust grains does not surpass the gravitational pull if the dust
opacity for the direct light
$350 \left(Z/Z_{\odot} \right) \rm{cm^{2} g^{-1}}$ is
smaller than the electron scattering opacity, i.e.,
\begin{eqnarray}
Z < Z_{\rm{limit}} = 1.0  \times 10^{-3}Z_{\odot}. \label{0124.1}
\end{eqnarray}
This critical metallicity becomes important in the cases with $10^{-3}M_{\odot} {\rm yr^{-1}}$,
where the formation of the HII region sets the upper mass limits
at $Z \lesssim Z_{\rm limit}$.
With the lower rates $\dot M \lesssim 10^{-4}M_{\odot} {\rm yr^{-1}}$, however, 
the HII region formation is more effective than the radiation pressure 
at $Z \simeq Z_{\rm limit}$, and limits the accretion at the lower stellar 
masses (Eq. \ref{0330.2.1}). 
As a result, the HII region formation sets the upper mass limits
also at higher metallicities, $Z \gtrsim Z_{\rm limit}$.

In what follows, we present in more detail our
numerical models of the accretion envelope fully coupled with
frequency-dependent radiation transport.
We derive the upper mass limits from such numerical modeling, and
compare them to the analytic estimates above.

\section{METHOD OF CALCULATIONS of accretion envelopes}
\label{sec.method_env}

 In this section, we describe how we construct the
numerical models of the accretion envelope, with which we investigate the
upper stellar mass limits for various metallicities $Z$ and accretion
rates $\dot M$ in Section \ref{sec.kekka_env}.
We determine the envelope structure consistently with
the stellar evolution calculations presented in Section \ref{stellar.evlol},
i.e., with the stellar luminosity $L_*$ and radius $R_*$
for a given stellar model with the mass $M_*$.
For simplicity, we assume the spherically symmetric and steady accretion flows.


 In the modeling of the accretion envelope, we construct
solutions where the radial distributions of the density and temperature
$(\rho , T)$ and those of the radiation energy density and
flux are consistent with each other.
The procedures are divided into the following two steps:
	\begin{enumerate}
\item For given radial profiles of the density and temperature $(\rho , T)$,
we calculate the radiative transfer to obtain the radiative energy-density
and flux distributions.
 \item We then update the density and temperature distributions
using the radiation field distribution obtained above.
	\end{enumerate}
We iterate the above two steps until we get numerical convergence,
i.e., the distributions $(\rho,T)$ no longer change anywhere in the envelope.

\subsection{Radiative transfer}\label{radiation_transfer}

\begin{figure}
	\begin{center}
		\includegraphics[width=\columnwidth]{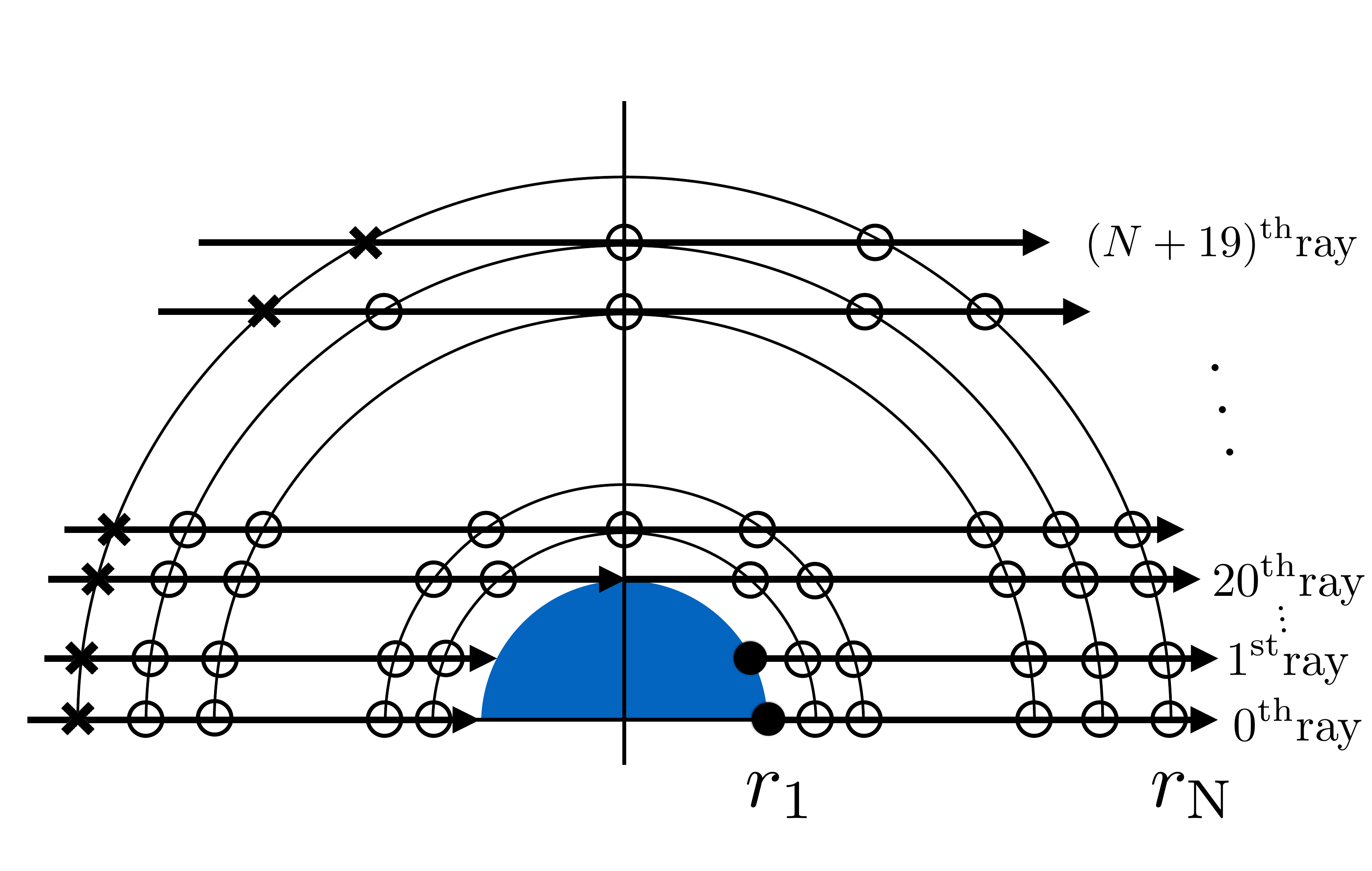}
	\end{center}

	\caption{ The geometry of rays for the ray-tracing calculation. 
The central filled half circle denotes the protostar, and the outer
concentric black circles represent the spherical grids, whose 
total number is $N=600$ for our default setting. The inner and outer boundaries
are at $r= r_{\rm 1}$ and $r_{\rm N}$ respectively.
The horizontal arrows represent the rays distributed over the computational
domain. We set 20 rays which penetrate the star. 
The inner boundary condition (Eq. \ref{1129.2.1}) is applied 
on the black circles on the stellar surface, while on the outer-boundary points (i.e., crosses) 
the condition of no-incoming intensity is imposed. }
	\label{fig0209.1}
\end{figure}

 We solve the frequency-dependent radiative transfer equation
using the variable Eddington factor method \citep[e.g.,][]{Mihalas_Mihalas1984}.
Our calculation covers the wavelength range
of $300~\AA \leq \lambda \leq 3~\rm{cm}$
($10^{10} \rm{Hz} \leq \nu \leq 10^{16} \rm{Hz}$ in frequency)
with 50 bins equally distributed on a logarithmic scale.


Figure \ref{fig0209.1} shows the geometry of rays along which
we solve the radiative transfer equation
\begin{eqnarray}
	\frac{d I_{\nu}}{d s} = - \rho \kappa_{\nu} \left( I _{\nu} - S_{\nu} \right),
\label{1218.2}
\end{eqnarray}
where $I_{\nu}(r,\mu)$ is the intensity, $s$ the distance along a ray, $\mu$ the cosine
between the ray and the line connecting a given point and the center, and $S_{\nu}$
the source function defined as
\begin{eqnarray}
	S_{\nu}
= \frac{ \kappa^{\rm abs}_{\nu} B_{\nu} + \sigma^{\rm sc} J_{\nu}}{\kappa^{\rm abs} + \sigma^{\rm sc}}, \label{1118.4}
\end{eqnarray}
where $\kappa^{\rm abs}_{\nu}$ and $\sigma^{\rm sc}_{\nu}$ are the
absorption and scattering opacity, and $J_{\nu}$ is the mean intensity.
We set 20 rays passing through the protostar, and also put rays
on the all spherical grids oriented in the tangential directions.
To solve equation \eqref{1218.2}, $B_{\nu}$ is obtained with a given
temperature distribution. 
With a guessed distribution of $J_{\nu}$. we can solve the ray-tracing equations 
\label{1218.2} with the following boundary conditions. 
For the inner boundary condition,
we assume the diffusion approximation
	\begin{eqnarray}
		I_{\nu} = B_{\nu} + \frac{\mu}{\rho \kappa_{\nu}} \frac{d B_{\nu}}{d r} = B_{\nu} + \frac{\mu}{\rho \kappa_{\nu}} \frac{\partial B_{\nu}}{\partial T} \frac{dT}{dr}, \label{1129.2.1}
	\end{eqnarray}
which is valid near the stellar surface except in a brief evolutionary
stage where the protostar swells and the optical depth decreases.
The temperature gradient in equation \eqref{1129.2.1} is given by
		\begin{eqnarray}
			\frac{dT}{dr} = 3 H_{\nu} \rho \kappa_{\nu} \left( \frac{\partial B_{\nu}}{\partial T} \right)^{-1}, \label{0105.1}
		\end{eqnarray}
where $H_{\nu} \equiv F_{\nu} / (4 \pi)$ is written as
		\begin{eqnarray}
			H_{\nu} = \frac{L_{*}}{(4 \pi R_{*})^2} \left( \frac{\partial B_{\nu}}{\partial T} \right) \bigg{/} \left( \int ^{\infty}_{0} \frac{\partial B_{\nu}}{\partial T} d \nu \right). \label{1129.2.2}
		\end{eqnarray}
As for the outer boundary condition, we assume that no radiation
comes into the computational domain.
We calculate the Eddington factor with the intensity $I_{\nu}(r,\mu)$
obtained by the ray-tracing,
\begin{eqnarray}
  f_{\nu}(r) = K_{\nu} / J_{\nu}
 = \left( \int ^{1}_{-1} I_{\nu} (r,\mu) \mu ^{2} d \mu \right) \bigg{/} \left( \int ^{1}_{-1} I_{\nu} (r,\mu) d \mu  \right), \label{0117.1}
\end{eqnarray}
with which we integrate the moment equations of the radiation transfer
equation \eqref{1218.2},
	\begin{eqnarray}
			\frac{1}{r^2} \frac{\partial \left( r^2 H_{\nu} \right)}{\partial r} = - \kappa_{\nu} \rho \left( J_{\nu} - S_{\nu} \right) \label{1118.1} \\
			\frac{\partial \left( f_{\nu} J_{\nu} \right)}{\partial r} + \frac{\left( 3 f_{\nu} - 1 \right) J_{\nu}}{r} = - \kappa_{\nu} \rho H_{\nu} ,\label{1118.2}
		\end{eqnarray}
where $J_{\nu}$ and $H_{\nu}$ are the zeroth and the first moment of intensity.
We construct the solution which satisfies the boundary conditions with a small number
of iterations (typically dozens), by transforming the term $\kappa_{\nu} B_{\nu}$ in the source function
$S_{\nu}$ (Eq. \ref{1118.4}) into the form only including $J_{\nu}$
(see Appendix~\ref{apdi1} for more details).

\begin{figure}
	\begin{center}
		\includegraphics[width=\columnwidth]{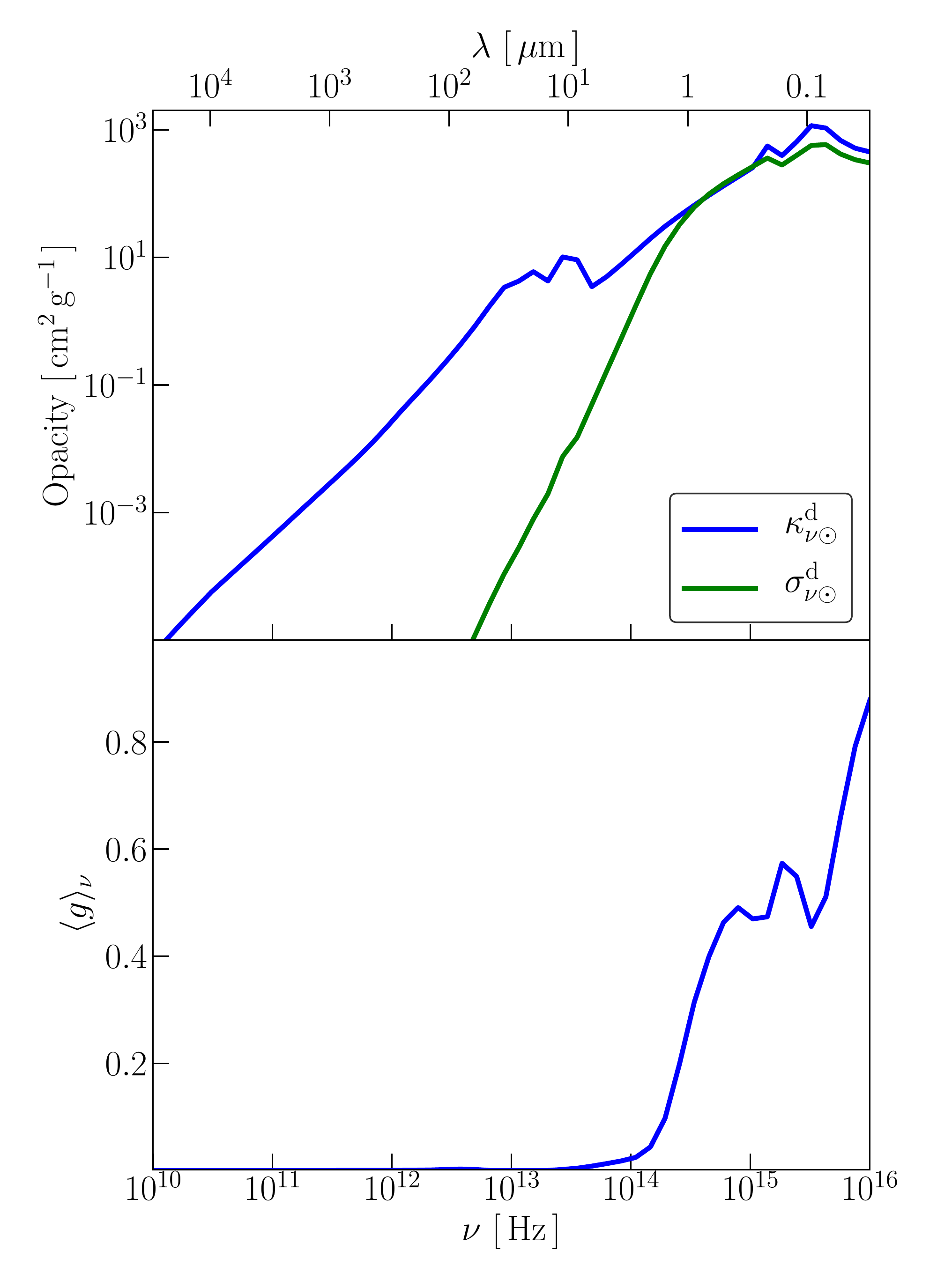}
	\end{center}
	\caption{The grain opacities used in our calculation for the accretion envelope. Top panel: the frequency dependence of the absorption opacity $\kappa_{\nu \odot}^{\rm {d}}$ and the scattering opacity $\sigma_{\nu \odot}^{\rm d}$. Bottom panel: same as the top panel but for $\langle g \rangle _{\nu}$}
	\label{fig1.1}
\end{figure}

Figure \ref{fig1.1} shows the frequency-dependent dust absorption
and scattering opacity we adopt.
Since the dust opacity has the strong frequency-dependence,
the radiation pressure on the dust cocoon cannot be evaluated accurately
without frequency-dependent radiative transfer calculations.
\citep{Wolfire_Cassinelli1986,Edgar_Clarke2003}.
We assume that the dust opacity is simply in proportion to the
metallicity,
	\begin{eqnarray}
			\kappa_{\nu}^{\rm d} = \left( \frac{Z}{Z_{\odot}} \right) \kappa_{\nu \odot}^{\rm d} \label{0209.1}
	\end{eqnarray}
	\begin{eqnarray}
			\sigma_{\nu}^{\rm d} = \left( \frac{Z}{Z_{\odot}} \right) (1 -  \langle g \rangle _{\nu}  ) \sigma_{\nu \odot}^{\rm d}, \label{0209.2}
	\end{eqnarray}
where $\langle g \rangle_{\nu}$ is the rate of forward scattering. 
We use the interstellar values for $\kappa_{\nu \odot}^{\rm d}$,$\sigma_{\nu \odot}^{\rm d}$ and $\langle g \rangle_{\nu}$ as in
\citet{Wolfire_Cassinelli1986}, also assuming the
two components of graphite and silicate grains.
The dust cross sections are calculated by means of the Mie theory
\citep{Bohren_Huffman1983,Wolfire_Cassinelli1986},
using the dielectric functions given by \citet{Draine_Lee1984} and \citet{Draine2003b}.
We assume the so-called MRN mixture \citep{MRN} for the grain size distribution,
	\begin{eqnarray}
				n_{ i} (a) da = C_{{i}} a^{-3.5}da ,(i = {\rm{graphite, silicate}}),  \label{eq1.11}
	\end{eqnarray}
where $C_{ i}$ is the scale factor also by \citet{Draine_Lee1984}.
The maximum and minimum grain sizes are $0.25\rm{\mu m}$
and $0.005\rm{\mu m}$ respectively.

The grains evaporate above the sublimation temperature
\begin{eqnarray}
	T_{\rm vap} = g \rho ^{\beta}, \label{eq1.12}
\end{eqnarray}
where $g = 2000$, and $\beta = 0.0195$ \citep{Isella_Natta2005}.
Since $\beta$ is small, the sublimation temperature depends only weakly
on the density and typically $T_{\rm vap} \simeq 1200$~K.
We incorporate this effect by multiplying the opacity by a
reduction factor $\epsilon$,
\begin{eqnarray}
 \epsilon(T, \rho) = \frac{1}{1 + \exp ((T - T_{\rm vap})/50)}, \label{eq1.14}
\end{eqnarray}
following \citet{Kuiper2010}.
Since we consider the gas opacity only for the absorption,
we use
	\begin{eqnarray}
		\kappa_{\nu}^{\rm abs} = \kappa^{\rm g} + \epsilon(T,\rho) \kappa^{\rm d}_{\nu}, \label{1218.3}
	\end{eqnarray}
where $\kappa^{\rm g}$ is the gas opacity, and for the scattering
	\begin{eqnarray}
		\sigma_{\nu}^{\rm sc} = \epsilon(T,\rho)  \sigma_{\nu}^{\rm d}.  \label{0117.2}
	\end{eqnarray}
We use the grey gas opacity because it does not strongly depend on frequency
in the relevant range in the stellar photosphere where the gas component dominates the opacity.
We use the same opacity tables as in \citet{Hosokawa_Omukai2009a}:
the OPAL tables for $T > 7000$~K \citep{Igresias1996}, and
the tables given by \citet{Alexander1994} for $T < 7000$~K
with modifications of removing the contribution from the dust grains
\citep{Asplund2005,Cunha2006}.

\subsection{Calculation of accretion flow  structures}

Next, we describe how to solve the flow structure, i.e., the radial distributions of
density and temperature for the radiation field obtained by solving
the radiative transfer equations (Sec.~\ref{radiation_transfer}).
Under the assumption of the spherical symmetry, the equation of motion is
\begin{eqnarray}
	u \frac{d u}{d r} = - \frac{G M_{*}}{r^2} + \frac{1}{c} \int  \kappa_{\nu} F_{\nu}  d \nu , \label{eq1.5}
\end{eqnarray}
where we ignore the term of the thermal pressure gradient
which is sufficiently smaller than the radiation pressure and gravity terms
\citep{Wolfire_Cassinelli1987}.
With a constant accretion rate, the velocity distribution $u(r)$ is translated
into the density distribution using the continuity equation \eqref{eq1.6.1}.
The energy equation is
	\begin{eqnarray}
		u \frac{d e}{dr} + P u \frac{d}{dr} \left( \frac{1}{\rho} \right) = - \Lambda_{\rm chem} - \Lambda_{\rm{rad}} ,\label{eq1.6}
	\end{eqnarray}
where $e=kT/\mu m_{\rm H} (\gamma - 1)$ is the internal energy per unit mass,
and $\gamma$ is the adiabatic exponent.
We also assume the equation of state
$P / \rho = (k T) / (\mu m_{\rm H}) $ for the ideal gas.
The radiative cooling rate $\Lambda_{\rm rad}$ is given by
	\begin{eqnarray}
		\Lambda_{\rm rad} = \Lambda_{\rm con} + \Lambda_{\rm line}, \label{0112.1}
	\end{eqnarray}
where $\Lambda_{\rm con}$ and $\Lambda_{\rm line}$
represent the contributions by the continuum and H$_2$ line emission,
\begin{eqnarray}
  \Lambda_{\rm con} =  4 \pi \left(
\int \kappa_{\nu}^{\rm abs} B_{\nu} d\nu  - \int \kappa_{\rm \nu}^{\rm abs} J_{\nu} d \nu
\right), \label{1117.2}
\end{eqnarray}
where $\kappa_{\nu}^{\rm abs}$ is the absorption opacity, and
\begin{eqnarray}
  \Lambda_{\rm line} =  \overline {\beta}_{\rm esc}
\Lambda_{\rm thin,H_{2}} e ^{\rm - \tau_{\rm C}} , \label{0113.2}
\end{eqnarray}
where $\Lambda_{\rm thin,H_{2}}$ is the cooling rate via optically
thin H$_2$ emission, $\overline {\beta}_{\rm esc}$
the line-averaged escape probability, and
$\tau_{\rm C}$ the continuum optical depth for the line emission.
We approximate $\Lambda_{\rm thin, H_{2}}$ using a fitting formula
given by \citet{Glover2015}.
The escape probability $\overline {\beta}_{\rm esc}$ is given as
a function of the column density of hydrogen molecules
	\begin{eqnarray}
		N_{\rm H_{2}} (r) = \int _{r}^{r^{'}} n({\rm H_{2}}) dr, \label{0405.1}
	\end{eqnarray}
where $n({\rm H_{2}})$ is the number density of hydrogen molecules
and $r^{'}$ is the outermost radius within which the Doppler shift
caused by the flow velocity is less than the line width,
$|v(r)-v(r^{'})|<(v_{\rm D}(r) + v_{\rm D}(r^{'}))/2$
(also see Appendix~\ref{apdi3} for more details).
The line cooling with other molecules such as
$\rm{CO}$ and $\rm{H_{\rm 2}O}$ is ignored because its contribution is minor in comparison
to the dust continuum cooling with the non-zero metallicities.
In our models, in fact, the H$_2$ line cooling is only effective in the
primordial case.
We calculate the chemical cooling rate $\Lambda_{\rm chem}$ as
\begin{eqnarray}
		\Lambda_{\rm chem} = - u \frac{d \epsilon_{\rm chem}}{dr}, \label{1117.3}
\end{eqnarray}
where $\epsilon_{\rm chem}$ is chemical binding energy.
We consider the species of $\rm{H_{2},H,H^{+},He,He^{+},He^{2+}}$ and $\rm{e^{-}}$,
whose abundances are obtained by solving the Saha equations.


We set the outer boundary at $r = r_{\rm out}$, where the density is 
$\rho _{\rm out} = 10^{-19} \rm{g~cm^{-3}}$ assuming the
free-fall profile \citep{Wolfire_Cassinelli1987},
	\begin{eqnarray}
		r_{\rm out} = \left( \frac{\dot M}{4 \pi \rho_{\rm out} \sqrt{2 G M_{\rm *}}} \right)^{2/3} . \label{eq1.8}
	\end{eqnarray}
The velocity at $r = r_{\rm out}$ is given as
$u_{\rm out} = \sqrt{{2 G_{\rm eff} M_{*}}/{r_{\rm out}}}$,
 where $G_{\rm eff}$ is the effective gravitational constant
considering the radiation-pressure effect,
\begin{eqnarray}
  G_{\rm eff} = G
\left( 1 -   \frac{\int \kappa_{\nu} L_{\nu} d \nu}{4 \pi G M_{*} c} \right), \label{1218.1}
\end{eqnarray}
where $\kappa_{\nu} = \kappa_\nu^{\rm abs} + \sigma_\nu^{\rm sc}$.
The temperature at the outer boundary is determined by the
thermal balance between the compression heating and radiative cooling
	\begin{eqnarray}
		P u_{\rm out} \frac{d}{dr} \left( \frac{1}{\rho_{\rm out}} \right) = -\Lambda_{\rm con} - \Lambda_{\rm line}.\label{eq1.9}
	\end{eqnarray}
Since the temperature in the dust cocoon is mostly determined by the 
local thermal balance between the absorption and emission of the continuum emission, 
i.e., $\Lambda_{\rm con} = 0$ in Equation \eqref{1117.2}, structure of the dust 
cocoon is actually insensitive to the outer boundary condition of the temperature.

\section{Results for cases with constant accretion rates}
\label{sec.kekka_env}

\begin{figure}
	\begin{center}
		\includegraphics[width=\columnwidth]{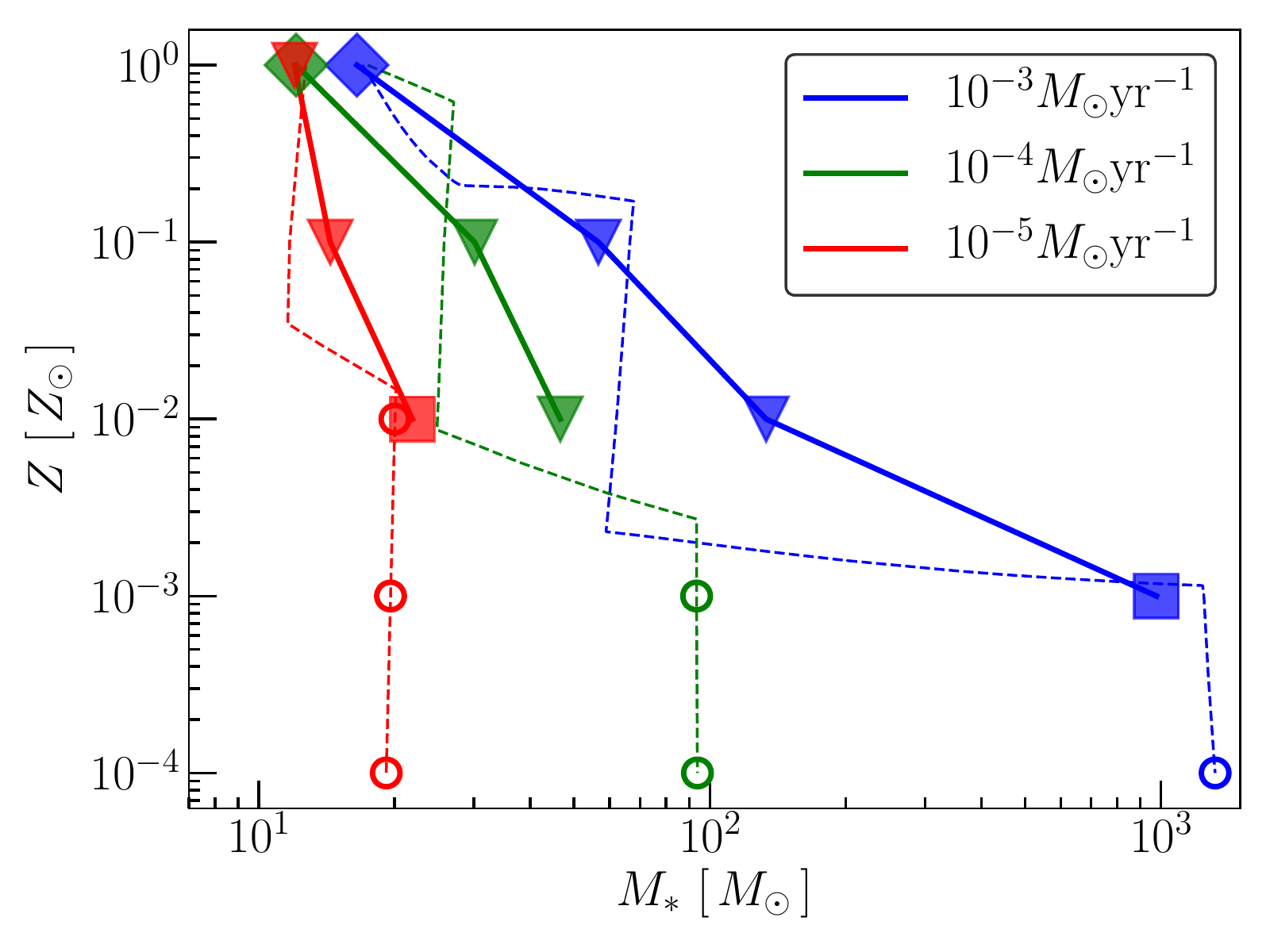}
	\end{center}
	\caption{The protostellar upper mass limits with different metallicities
at constant mass accretion rates of $\dot M = 10^{-3}$,$10^{-4}$,
and $10^{-5}M_{\odot} \rm{yr^{-1}}$.
The symbols indicate the optical depth of the cocoon and 
dominant mechanisms for halting the inflow at the limits:
(i) the cocoon is optically thick to both the diffuse and direct radiation and the accretion is terminated by radiation pressure by the diffuse light ($\Diamond$);
(ii) the cocoon is optically thin to the diffuse radiation but still thick to the direct radiation, 
and the flow is decelerated mostly by the direct radiation pressure around the dust destruction front
($\triangledown$);
(iii) the cocoon is optically thin also to diffuse radiation, and so the direct radiation pressure onto the entire cocoon 
decelerates the flow ($\Box$).
The dashed lines represent the upper mass limits by the analytic argument as in Figure \ref{fig0124.1}.
At lowest metallicities, the radiation pressure on to the dust cocoon is not effective and the
stellar mass is limited by the growth of the HII region. 
In such cases, we show those limits by the symbol $\bigcirc$ from the analytic argument.
}
	\label{fig0127.1}
\end{figure}

\begin{figure*}
	\begin{center}
		\includegraphics[width=165mm]{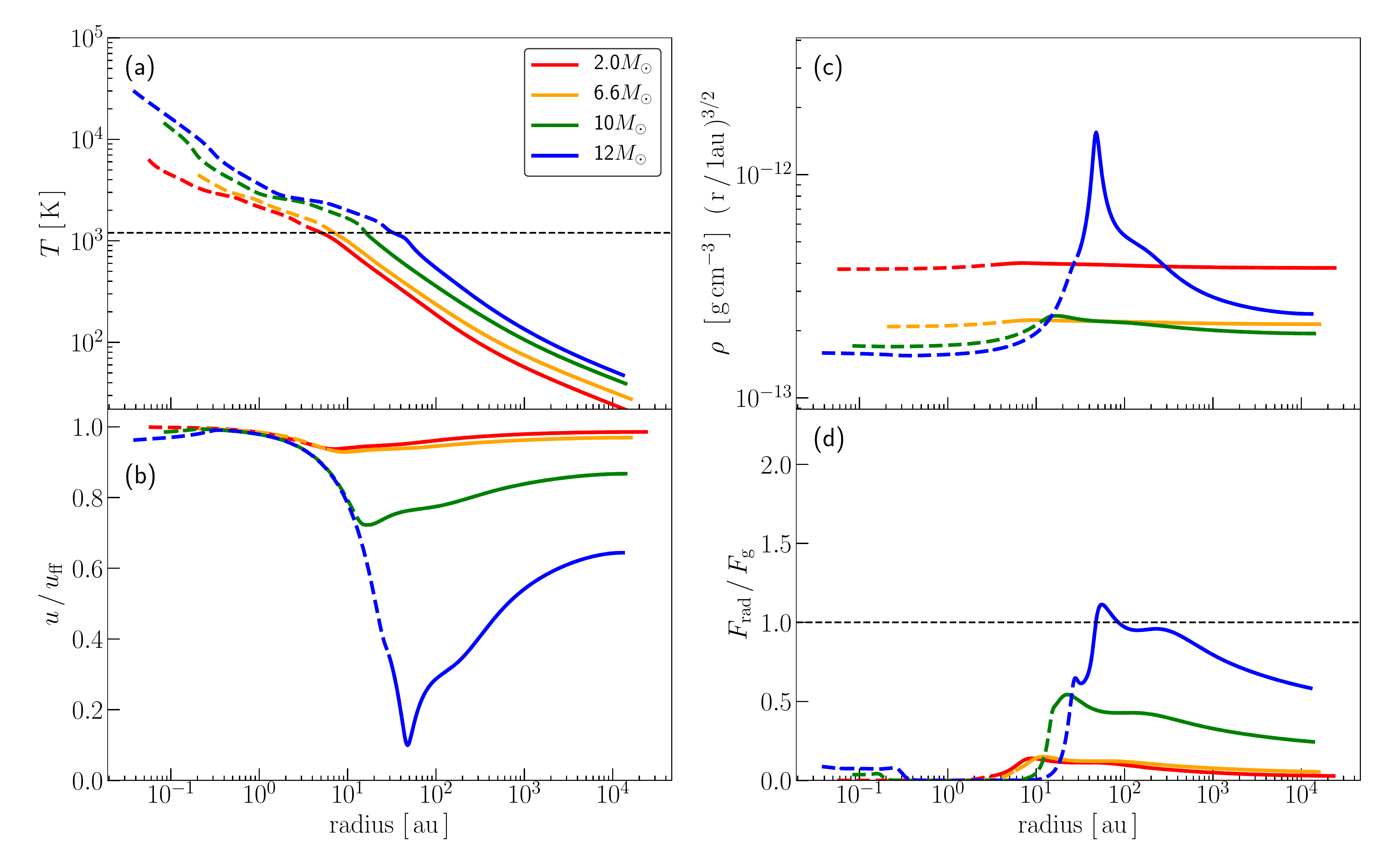}
	\end{center}
	\caption{The evolution of the envelope structure with metallicity $Z = 1Z_{\odot}$ and accretion rate $\dot M = 10^{-4}M_{\odot} \rm{yr^{-1}}$. 
In each panel, the solid and dashed lines indicate the regions where the dust is present and absent, 
respectively. 
Panel (a): temperature distributions. The thin dashed horizontal line indicates the dust sublimation temperature of $1200\rm{K}$. 
(b): velocity distributions normalized by the free-fall value. 
(c): The density distributions. The density is multiplied by $r^{1.5}$ to emphasize the deviations from the free-fall density profile. 
(d): The distribution of the ratio of the radiation pressure to the gravity. 
In all of the panels, the different colors represent the different epochs which correspond 
to the following different stages of the protostellar evolution: 
the adiabatic accretion (at $2.0M_{\odot}$), swelling (at $6.6M_{\odot}$) and KH contraction (at 
$10$ and $12M_{\odot}$). In panel (b), we can see that the minimum value of the velocity ratio 
$u/u_{\rm ff}$ becomes 1/10 at the epoch of $M_* = 12M_{\odot}$. 
The inner end of each line is the radius of the protostar.}
	\label{fig0126.1}
\end{figure*}

\begin{figure}
	\begin{center}
		\includegraphics[width=\columnwidth]{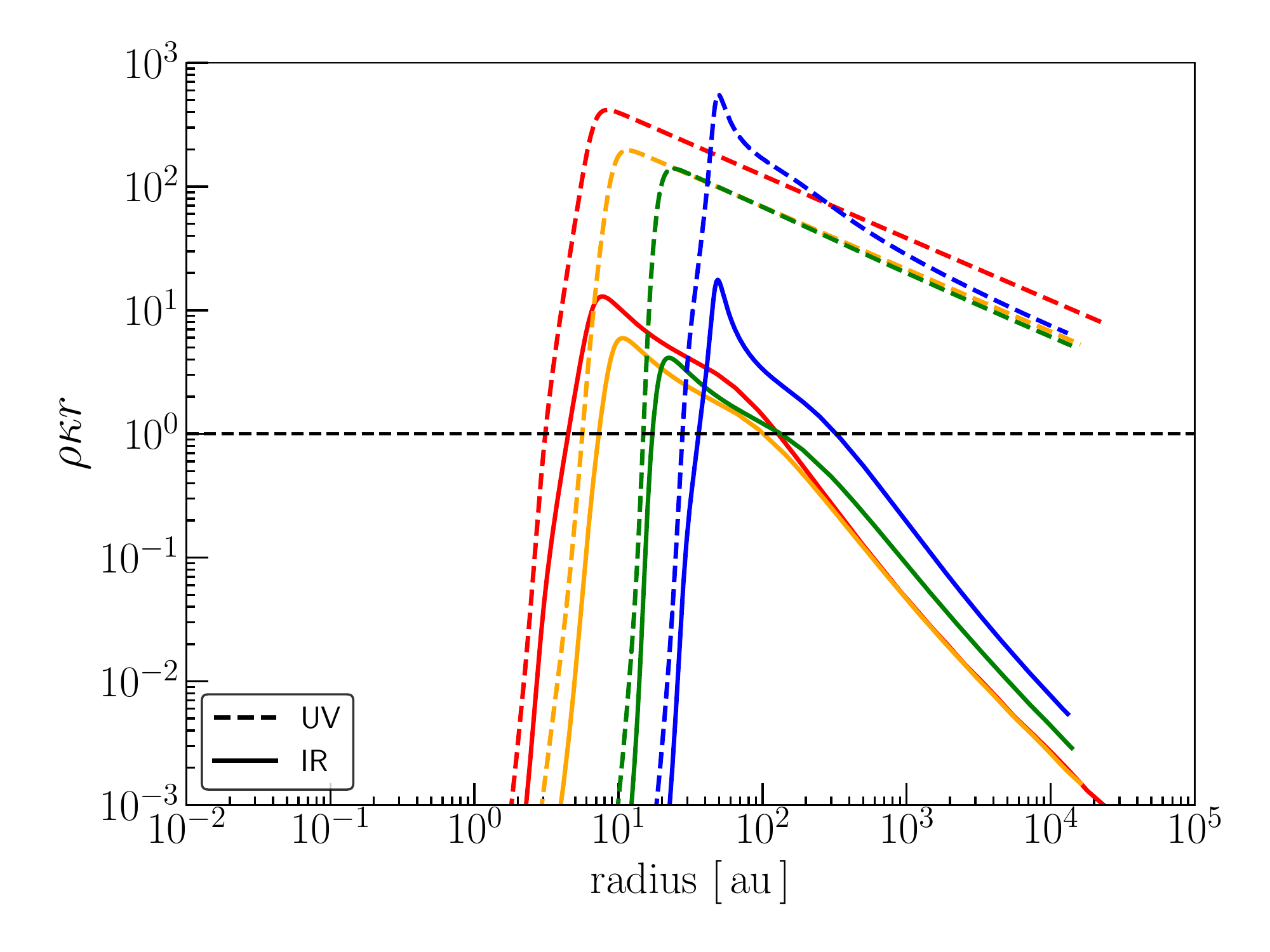}
	\end{center}
	\caption{Radial distribution of local optical depth for the case of $Z = 1Z_{\odot}$ 
and $\dot M = 10^{-4}M_{\odot} \rm{yr^{-1}}$. The solid and dashed lines shows those for 
the IR and UV range, $\rho \kappa_{\rm IR} r$ and $\rho \kappa_{\rm UV} r$. 
The same colors indicate the same epochs shown in Figure \ref{fig0126.1}. }
	\label{fig0126.3}
\end{figure}

\begin{figure}
	\begin{center}
		\includegraphics[width=\columnwidth]{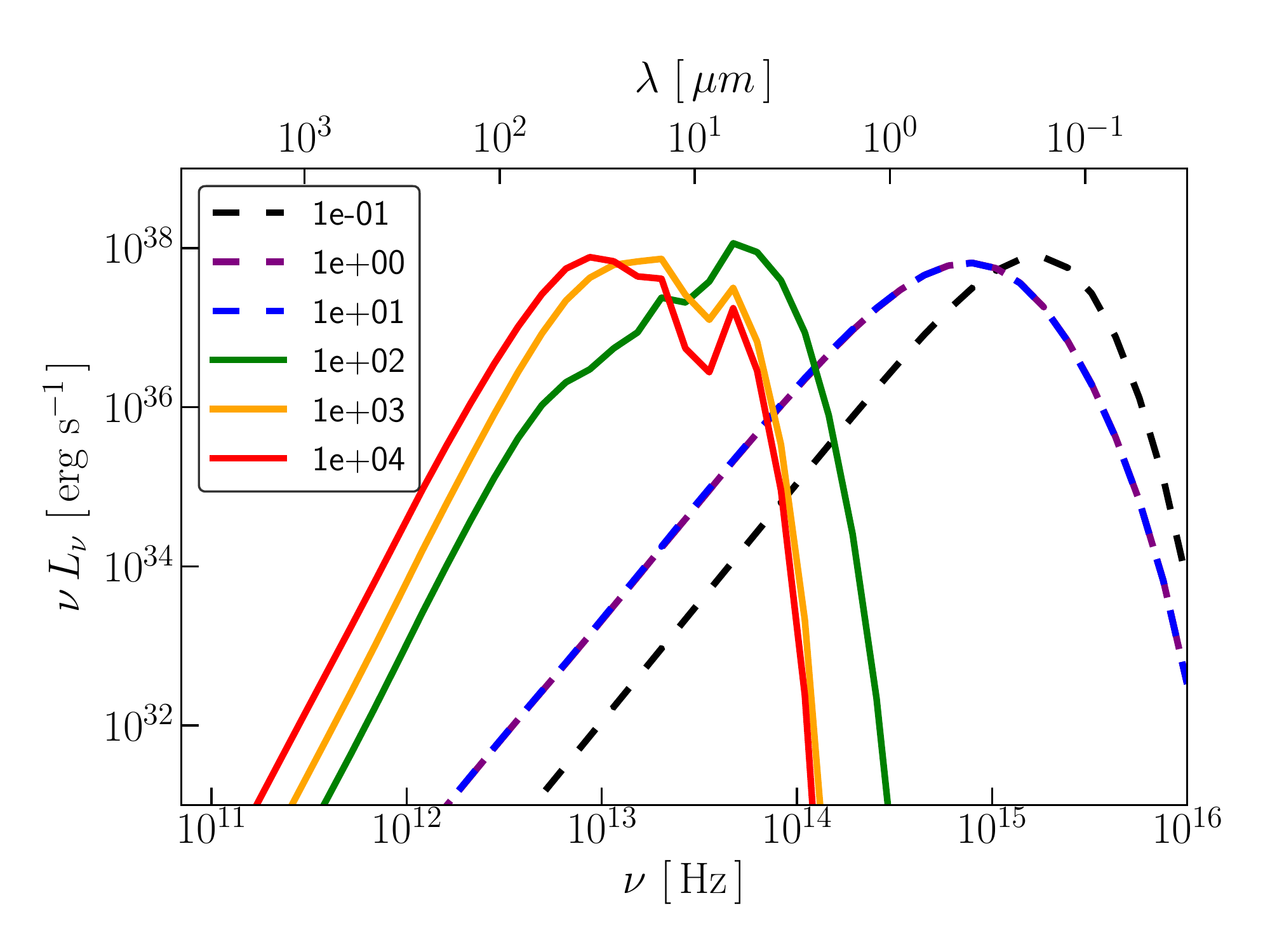}
	\end{center}
	\caption{The spectral energy distributions at different radii 
in the envelope at the epoch of $M_* = 12 M_\odot$ for the case with $Z = 1Z_{\odot}$ and 
$\dot M = 10^{-4}M_{\odot} \rm{yr^{-1}}$. 
The different lines show those at the radii $r = 0.1 - 10^{4} \rm{au}$ indicated in the legend. 
The solid and dashed parts of the lines correspond to the region where dust is present and absent, 
respectively.}
	\label{fig0126.2}
\end{figure}

We here see the evolution of envelope structures with constant accretion 
rates $\dot M=10^{-3}$, $10^{-4}$, $10^{-5}M_{\odot} \rm {yr^{-1}}$, 
and estimate the upper mass limits. 
We consider the variation of metallicities covering $Z=1$, $10^{-1}$,
$10^{-2}$, $10^{-3}$, and $10^{-4}Z_{\odot}$ for each accretion rate.
Since the radiation pressure exerted on dust grains is not efficient
for terminating the accretion flow at 
$\la 10^{-3}Z_{\odot}$ (Eq.~\ref{0124.1}), we only consider
the cases with $Z \geq 10^{-4}Z_{\odot}$ in our numerical calculations. 
Note that only the effect of the radiation pressure 
on the dust cocoon is consistently taken into account here.
The effect of the feedback caused by an expanding HII region 
just follows our analytic consideration presented in Section \ref{sec.kaiseki}.


Before examining the numerical results in more detail,
we overview the resulting upper mass limits for the various cases 
in Figure \ref{fig0127.1} (marked by the symbols $\Diamond$, $\triangledown$, and $\Box$).
At lowest metallicities where the upper mass limits are set by the growth of HII regions,
the limits by the analytic argument are indicated by circles $\bigcirc$.
The analytical estimates given in Section \ref{sec.kaiseki} are also 
shown by the dashed lines for comparison.
We see quite good agreements between
the numerical and analytical results.
The dominant feedback mechanisms are identical for almost all
of the examined parameter space. 
Some disagreement appears only for the
cases where the envelope is optically thick for the direct light 
but optically thin for the diffuse light (marked by the symbols $\triangledown$):
the analytic argument tells that the upper limit should not depend on metallicity,
while it actually increases with decreasing metallicity 
according to our numerical results.

Below we see the evolution of envelope structures in the order of 
decreasing the optical depth of the dust cocoon, where the radiation pressure
operates in different ways to terminate the accretion flow.

\subsection{Cases that the dust cocoon is optically thick to the diffuse light}
\label{IR_thick}

We begin with the case of $Z=1Z_{\odot}$ and $\dot M=10^{-4}M_{\odot}\rm{yr^{-1}}$.
Our analytic model predicts that the accretion terminates at
$M_* \simeq 17M_{\odot}$ by the diffuse radiation pressure in this case,
where the envelope remains optically thick to the diffuse light 
throughout the accretion stage (see Section~\ref{sec.kaiseki}).

Figure \ref{fig0126.1} shows the distributions of the temperature, density, 
infall velocity, and the ratio of radiation pressure to the gravity force 
at different epochs in this case.
The infall velocity is normalized by the free-fall velocity.
The density distribution is plotted as $\rho r^{3/2}$ to emphasize the deviation from that in the free-fall case  (i.e., $\rho \propto r^{-3/2}$).
The solid part of the lines corresponds to the region where the dust is present,
while the dashed part to the region where it has already been evaporated.
Here, we consider the dust destruction radius as the point where 
the dust reduction factor $\epsilon$ defined by Equation \eqref{eq1.14} 
takes the value of 0.1.
The four lines in each panel correspond to the different stages of the protostellar evolution:
the adiabatic accretion ($2M_{\odot}$), swelling ($6.6M_{\odot}$) 
and KH contraction phases ($10M_{\odot},12M_{\odot}$).
Also at the final snapshot for $M_* \simeq 12M_{\odot}$, 
the accretion flow is being stopped by the radiation pressure.

Temperature is relatively low with $<100\rm{K}$ around the outer boundary, 
but increases in the inner region.
It reaches the dust sublimation value of $1200\rm{K}$ at radius $10-100 \rm{au}$, 
depending on the protostellar mass.
The temperature continues to rise inward also in the dust-free region up to the stellar surface.
With increasing the protostellar mass and so the luminosity, the temperature is higher 
at a fixed radius because of the higher radiative heating.
Note that the dust destruction front moves outwards with increasing the 
stellar mass since the dust sublimation temperature is constant at $1200\rm{K}$.

Radiation pressure to the accretion flow increases with growing stellar luminosity.
While the protostar is still small ($2.0$ and $6.6M_{\odot}$),
the flow is close to the free fall (Figure \ref{fig0126.1} b and d) due to negligible radiation pressure.
Once the stellar mass exceeds $\sim 10M_{\odot}$, the radiation pressure 
effect becomes remarkable. Especially, when the stellar mass reaches $12M_{\odot}$, 
radiation pressure almost balances with the gravity within $300 \rm{au}$,
and eventually exceeds it around $100 \rm{au}$.
As a consequence, the flow is decelerated to 10 \% of the free-fall velocity 
around $50 \rm{au}$.
The slow-down of the flow also affects the density distribution.
As seen in Figure \ref{fig0126.1} b, the density follows the free-fall 
law of $\propto r^{-3/2}$ at early epochs with $M_* \simeq 2.0~M_\odot$ and
$6.6~M_\odot$.
At $M_* \simeq 12M_{\odot}$, however, the radiation pressure effect becomes conspicuous 
and a shell-like structure is formed at $\simeq 50 \rm{au}$, 
where the radiation pressure effect is the most prominent.

Recall that, in Section \ref{stellar.evlol}, we have assumed the free-fall
flow at the photosphere as the outer boundary condition for 
our numerical models of the accreting protostars.
We note that this is always justified even if
the accretion flow is greatly decelerated through the dust cocoon. 
For example, at the epoch of $M_* \simeq 12~M_\odot$ where the flow is remarkably 
decelerated at $\simeq 50$~au, the flow accelerates again after passing 
through the dust destruction front owing to very small gas opacity.
As a result, the velocity recovers the free-fall value before reaching 
$\sim 0.1~\rm{au}$. Our modeling of the accreting protostar 
and surrounding dust cocoon is thus consistent each other. 


Figure \ref{fig0126.3} shows the radial distributions of the local optical 
depths for direct and diffuse light $\rho \kappa_{\rm UV} r $ 
and $\rho \kappa_{\rm IR} r $, for the same epochs as in Figure \ref{fig0126.1}.
Here, the opacity for the direct light $\kappa_{\rm UV}$ is estimated 
in the same way as in Section \ref{sec.optical_depth}
and that for the diffuse light is given by the Rosseland mean opacity at each radius.
The local optical depth is the ratio of the photon mean free path to the local physical scale,
and also represents the contribution to the optical depth $\tau=\int \rho \kappa dr$
from each region on the logarithmic scale as $\rho \kappa r = d \tau / d \ln r$.
The cocoon is optically thick both to the direct and diffuse light 
near the dust destruction front, and the peak values of the local 
optical depths are $\rho \kappa_{\rm UV} r \sim 10^2$ and $\rho \kappa_{\rm IR} r \sim 10$.
Except the final epoch of $M_* \simeq 12M_{\odot}$,
the peak values of $\rho \kappa_{\rm UV} r $ and $\rho \kappa_{\rm IR} r$
gradually decrease as the stellar mass increases, 
because the density decreases and the dust destruction front moves outward.
In contrast, the peak becomes higher for $M_* \simeq 12M_{\odot}$ 
than in the previous epochs since the density around the dust destruction 
front sharply increases due to the deceleration of the accretion flow.


Next we show in Figure \ref{fig0126.2} the radiative energy distribution 
inside the envelope at the different radii
$0.1,1,10,10^{2},10^{3}$ and $10^{4} \rm{au}$ at the epoch of 
$M_* \simeq 12~M_{\odot}$, when the radiation pressure almost halts the accretion flow. 
The dust destruction front locates at $20 \rm{au}$ in this moment, and the solid (dashed) lines
in this figure indicates those in the region where the dust is present (absent), respectively.
Since the gas opacity is large in a hot environment near the protostar,
the gas photosphere appears at $0.2 \rm{au}$.
Inside the photosphere, the spectral peak shifts to the higher frequency 
as the measuring point approaches the star because the temperature rises.
For example, the peak shifts from $0.3 \rm{\mu m}$ ($10^{15}\rm{Hz}$)
at 1au, outside the gas photosphere, to 
$0.15 \rm{\mu m}$ ($2\times 10^{15} \rm{Hz}$) at $0.1\rm{au}$, 
inside the photosphere. 
Between the gas photosphere and the dust destruction front ($1 - 10 \rm{au}$),
the envelope is optically thin and the radiation field is dominated by the direct light from the
photosphere, which is the blackbody at $10^{4}\rm{K}$.
Around the dust destruction front, the envelope is very optically thick and
the direct light is absorbed and re-emitted by the dust grains.
The main radiation component outside the dust destruction front ($>20 \rm{au}$)
is the diffuse light, peaking at the infrared
$10^{13}-10^{14} \rm{Hz}$ ($3-30 \rm{\mu m}$).
Outside the dust destruction front,
the optical depth $\rho \kappa_{\rm IR} r$
exceeds unity even for the diffuse light up to $10^{3}\rm{au}$, 
and the absorption and re-emission repeatedly occurs.
The radiation pressure almost balances the gravity in this region, 
and the flow velocity is nearly constant.
We can see that the flow is mostly decelerated 
around $50 \rm{au}$ outside the dust destruction front,
where re-emitted diffuse light is the dominant radiation component.
Finally, observed from outside, the spectrum peaks around $10^{13} \rm{Hz}$,
corresponding to the dust photospheric temperature of $150 \rm{K}$.
The dip around $10 \mu{\rm m}$ comes from the presence of
the silicate feature in the absorption opacity in this wavelength range (Figure \ref{fig1.1}).
Owing to the large absorption opacity, only photons emitted from outer cold regions
can escape from the cocoon, so the radiation intensity becomes weak in this band.

\subsection{Cases that the envelope is optically thick to the direct light but
thin to the diffuse light}\label{dest_ruct}
\begin{figure*}
	\begin{center}
		\includegraphics[width=170mm]{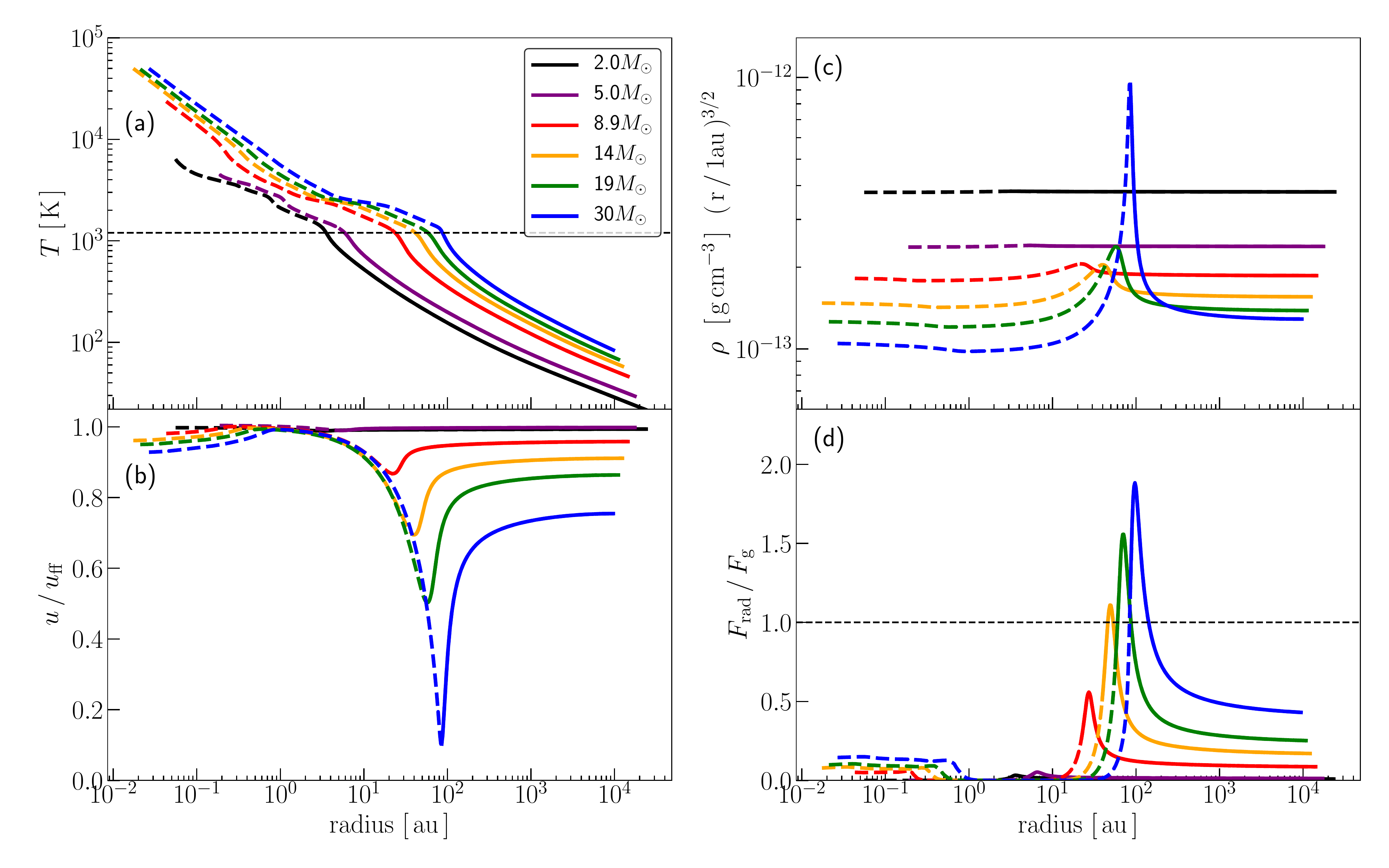}
	\end{center}
	\caption{Same as Figure \ref{fig0126.1}, but for the case of 
$Z = 10^{-1}Z_{\odot}$ and $\dot M = 10^{-4}M_{\odot} \rm{yr^{-1}}$.  
Shown are the distributions at the following stages: the adiabatic accretion ($2M_{\odot}$), 
swelling ($5M_{\odot}$), KH contraction ($8.9M_{\odot}$) and the main sequence accretion 
($14,20$ and $30M_{\odot}$). 
In panel (b), the minimum value of the velocity ratio $u/u_{\rm ff}$ is 
1/10 at the stellar mass of $30M_{\odot}$.}
	\label{fig0126.5}
\end{figure*}
\begin{figure}
	\begin{center}
		\includegraphics[width=\columnwidth]{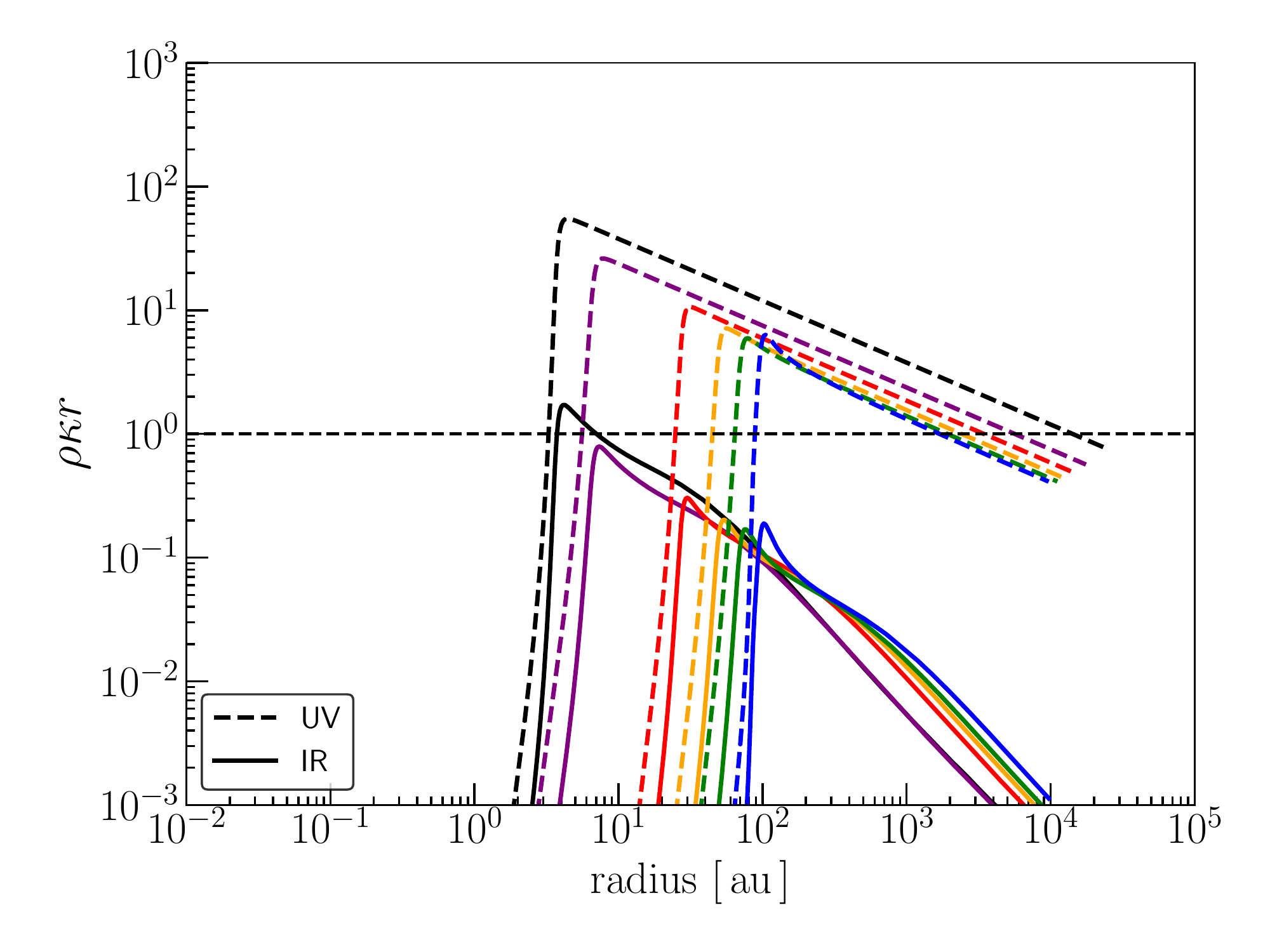}
	\end{center}
	\caption{Same as Figure \ref{fig0126.3}, but for the case with $Z = 10^{-1}Z_{\odot}$ and $\dot M = 10^{-4}M_{\odot} \rm{yr^{-1}}$. Shown with the same color are those at the same epochs 
as in Figure \ref{fig0126.5}.}
	\label{fig0126.7}
\end{figure}
\begin{figure}
	\begin{center}
		\includegraphics[width=\columnwidth]{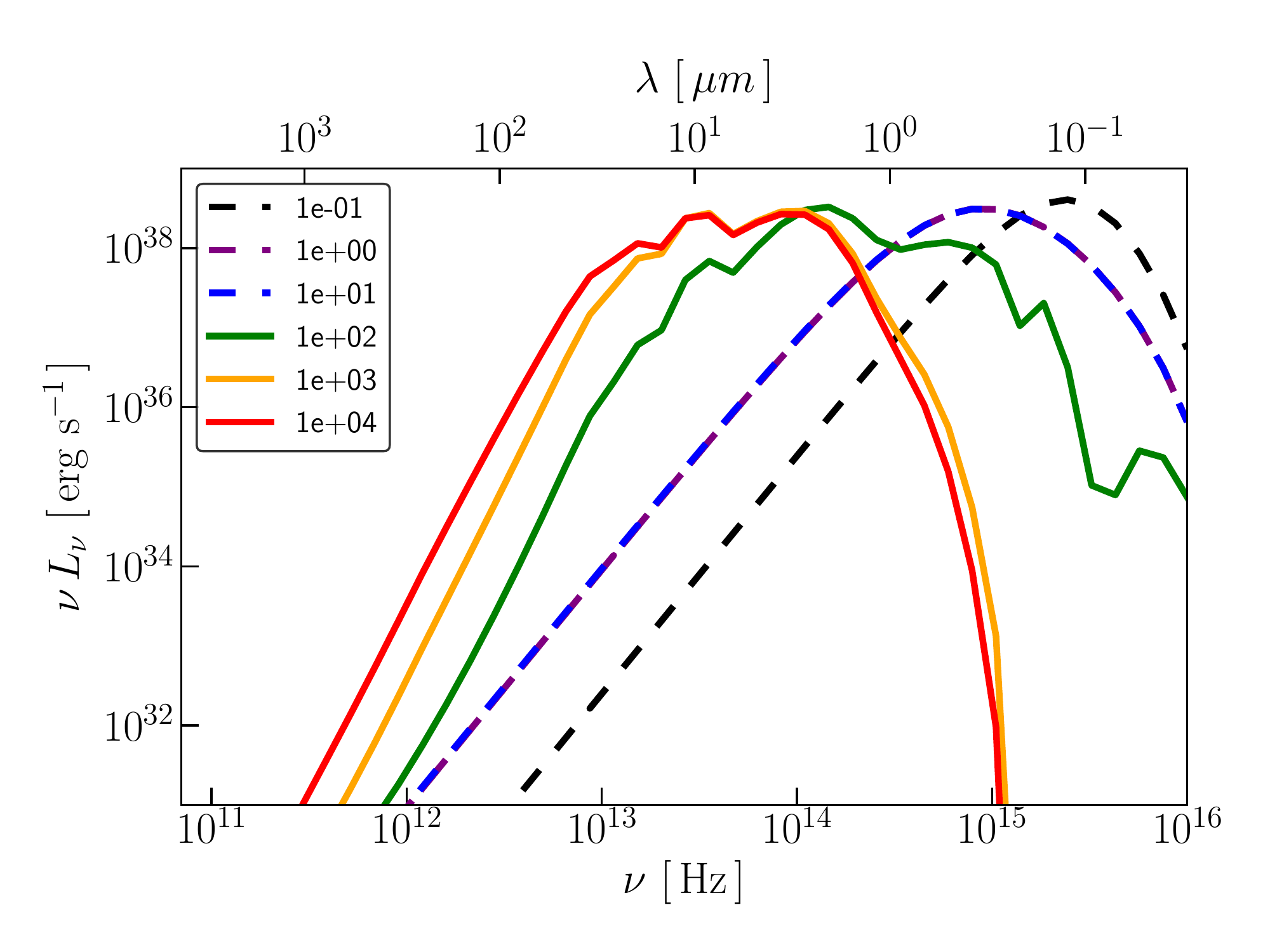}
	\end{center}
	\caption{Same as Figure \ref{fig0126.2}, but for the case with $Z = 10^{-1}Z_{\odot}$ and $\dot M = 10^{-4}M_{\odot} \rm{yr^{-1}}$ at the epoch of $30~M_\odot$, where the accretion is being halted. 
The lines are for those at the radii, $0.1\rm{AU}- 10^{4} \rm{au}$, indicated in the legend.}
	\label{fig0126.6}
\end{figure}

Next, we see the case of $Z = 10^{-1}Z_{\odot}$ and $\dot M = 10^{-4}M_{\odot} \rm{yr^{-1}}$ as an example.
According to the analysis in Section \ref{sec.kaiseki}, the dust cocoon is always optically thick
to the direct light while it becomes optically thin to the diffuse light for $\ga 5M_{\odot}$.
Also, it is predicted that the accretion flow is halted at $30M_{\odot}$
by radiation force by the direct light exerted on the dust destruction front
(Eq. \ref{0122.2} and Sec. \ref{dust_press}).

As in Figure \ref{fig0126.1}, Figure \ref{fig0126.5} shows the envelope
structure in this case at six different evolutionary stages: 
the adiabatic accretion ($2M_{\odot}$),
swelling ($5M_{\odot}$), KH contraction ($8.9M_{\odot}$), 
and main sequence accretion ($15,20$ and  $30M_{\odot}$) phases.
At the final epoch at $M_* \simeq 30M_{\odot}$, the accretion flow is 
being stopped by the radiation force.

As in the case seen in Section \ref{IR_thick}, the flow is almost the free fall
when the protostar is small ($2$ and  $5M_{\odot}$; Figure \ref{fig0126.5} b and d).
For $M_* \ga 9M_{\odot}$, the radiation force becomes as large as the gravitational attraction
near the dust destruction front.
When the protostellar mass exceeds $14M_{\odot}$, the radiation force exceeds
the gravity around $10^2{\rm au}$ and decelerates the flow.
At $30M_{\odot}$, the radiation force reaches about twice the gravity at $10^{2}\rm{au}$,
and the accretion flow is decelerated to 10\% of the free fall velocity.
Unlike the case in Section \ref{IR_thick} (at $12M_{\odot}$),
here the radiation force by the diffuse light, which is dominant outside 
a few 10 au, is only half of the gravity.
But more inside, the deceleration by the direct light sharply occurs in a narrow region
around the dust destruction front.
As a result, the density rises more steeply,
and a thinner shell-like structure is formed than seen in Section \ref{IR_thick} (Figure \ref{fig0126.5} c).

The local optical depths to the direct ($\rho \kappa_{\rm UV} r$)
and diffuse light ($\rho \kappa_{\rm IR} r$) are shown in Figure \ref{fig0126.7}.
The peak values of both $\rho \kappa_{\rm UV} r$ and $\rho \kappa_{\rm IR} r$ decrease
with the protostellar mass, except for the final phase of $30M_{\odot}$,
where the density is enhanced around the dust destruction front
due to the deceleration.
The cocoon is always optically thick to the direct light, with
maximum $\rho \kappa_{\rm UV} r \simeq 10$.
To the diffuse light, on the other hand,
while the cocoon is marginally optically thick with $\rho \kappa_{\rm IR} r \sim 1$
at early stages ($2$ and $5M_{\odot}$), it becomes optically thin for $\ga 9M_{\odot}$.

Next, in Figure \ref{fig0126.6}, the radiative energy distribution is shown at different radii
($0.1,1,10,10^{2},10^{3}$, and $10^{4} \rm{au}$) at the protostellar mass of $30M_{\odot}$,
where the accretion flow is decelerated remarkably by the radiation force.
The dust destruction front is around $10^{2} \rm{au}$: the inner three spectra ($0.1,1,10 \rm{au}$)
are those inside the dust free region and the rest ($10^2,10^{3},10^{4} \rm{au}$) 
are in the dusty region.
In the inner optically thin region ($1,10 \rm{au}$),
the spectrum is of the black-body at $2 \times 10^{4} \rm{K}$ emitted from the gas photosphere,
peaking at $0.3 \mu \rm{m}$ ($10^{15} \rm{Hz}$).
This direct light is absorbed and re-emitted at the dust destruction front.
In the spectrum at $10^{2} \rm{au}$ we can see that the direct light at $\sim 0.3 \mu{\rm m}$
has been absorbed and re-emitted as the diffuse light.
Since the dust cocoon is optically thin to the diffuse light, photons re-emitted
at the dust destruction front escapes freely without absorption.
In the outer dusty region ($10^{3}$ and $10^{4} \rm{au}$),
the radiation field is dominated by that re-emitted component 
at the dust destruction front, which
peaks at $4\rm{\mu \rm{m}}$ ($7 \times 10^{13}\rm{Hz}$) corresponding
to the dust sublimation temperature $1200\rm{K}$.
Since the dust cocoon is optically thin to the diffuse light, 
the spectrum is hardly modified and
can be directly observed from outside.
Contrary to those shown in Figure \ref{fig0126.2},
a bump appears around $10 \mu{\rm m}$ in the spectrum.
This is because more radiation is re-emitted at this frequency range
due to the higher value of opacity (Figure \ref{fig1.1}).

\subsection{Cases that the envelope is optically thin also to the direct light} \label{opticall_thin_for_UV}
\begin{figure*}
	\begin{center}
		\includegraphics[width=170mm]{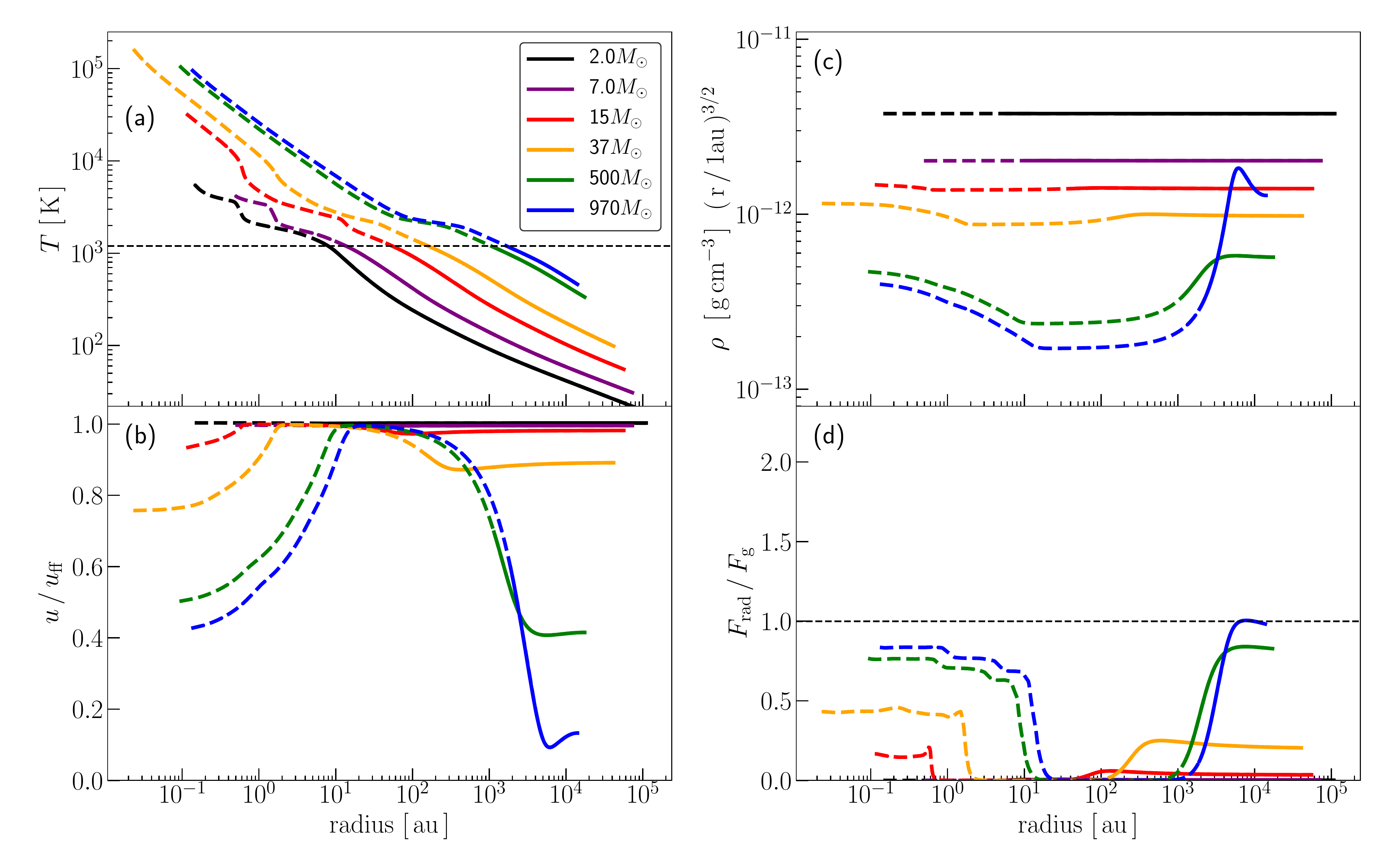}
	\end{center}
	\caption{Same as Figure \ref{fig0126.1}, 
but for $Z = 10^{-3}Z_{\odot}$ and $\dot M = 10^{-3}M_{\odot} \rm{yr^{-1}}$. 
Shown are those at the following evolutionary stage: the adiabatic accretion ($2M_{\odot}$), 
swelling ($7M_{\odot}$), KH contraction ($15M_{\odot}$) and 
the main-sequence accretion ($37, 510$ and $980M_{\odot}$). 
The minimum velocity ratio to the free-fall value is 1/10 at the epoch of $980M_{\odot}$.}
	\label{fig0126.9}
\end{figure*}

\begin{figure}
	\begin{center}
		\includegraphics[width=\columnwidth]{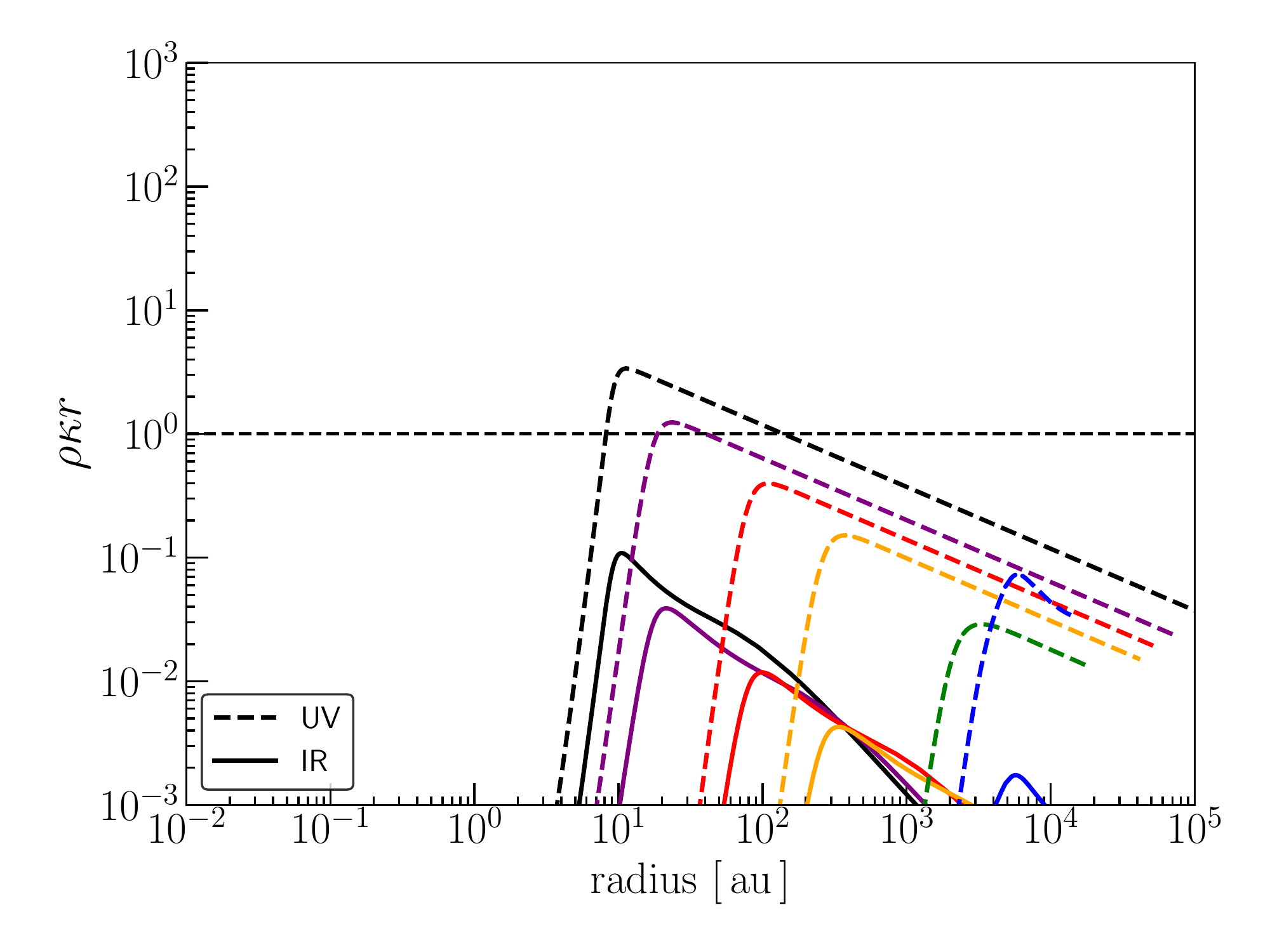}
	\end{center}
	\caption{Same as Figure \ref{fig0126.3}, but for the case with $Z = 10^{-3}Z_{\odot}$ and 
$\dot M = 10^{-3}M_{\odot} \rm{yr^{-1}}$. Shown with the same color are those at the same epochs 
as in Figure \ref{fig0126.9}. }
	\label{fig0126.11}
\end{figure}
\begin{figure}
	\begin{center}
		\includegraphics[width=\columnwidth]{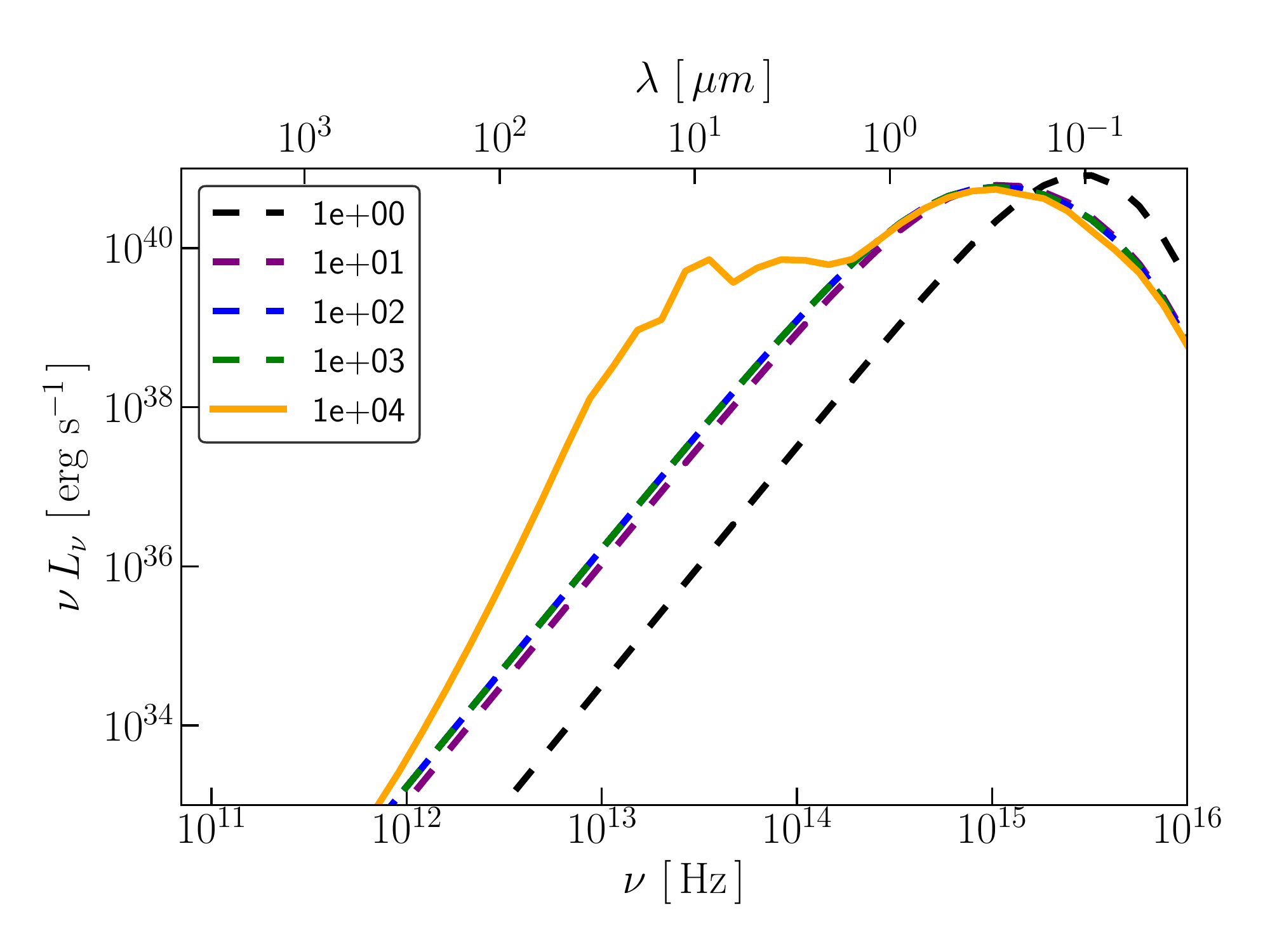}
	\end{center}
	\caption{Same as Figure \ref{fig0126.2}, but for the case with $Z = 10^{-3}Z_{\odot}$ and 
$\dot M = 10^{-3}M_{\odot} \rm{yr^{-1}}$ at the epoch of $980~M_\odot$, where the accretion is being halted. }
	\label{fig0126.10}
\end{figure}

Finally, we see the case of $Z = 10^{-3} Z_{\odot}$ and $\dot M = 10^{-3}M_{\odot} \rm{yr^{-1}}$.
According to the analytic argument in Section \ref{sec.kaiseki}, 
the dust cocoon is always optically thin
both to the diffuse and direct light.
The upper mass limit is predicted as $10^{3} M_{\odot}$ (Sec. \ref{dust_press})
due to radiation force by the direct light in the optically thin envelope
and also to the expansion of the HII region.
Note that the effect of the HII region formation is 
not included in the current numerical model but is estimated by
the analytic argument.

The evolution of the envelope structure is shown in Figure \ref{fig0126.9}, where
the six lines correspond to epochs of the adiabatic accretion ($2M_{\odot}$),
swelling ($7M_{\odot}$), KH contraction ($15M_{\odot}$) and main sequence accretion
($37,510$ and $980M_{\odot}$), respectively.
At $980M_{\odot}$, accretion flow is being stopped by the radiation pressure.

The radiation force becomes conspicuous when the protostar becomes more massive than $15M_{\odot}$.
Unlike the cases seen above in Sections \ref{IR_thick} and \ref{dest_ruct},
radiation pressure becomes strong near the inner and outer boundaries
(Figure \ref{fig0126.9} d).
Around the outer boundary, the dust grains work as the opacity source.
Since the envelope is very optically thin and the direct light is hardly attenuated,
the ratio of radiation pressure to gravity remains constant.
On the other hand, the opacity in the inner region is dominated by the electron scattering.
For $M_* \ga 15M_{\odot}$, although the photoionization is not taken 
into account in this calculation,
thermal ionization proceeds in the region where the temperature exceeds $10^{4} \rm{K}$
near the star (Figure \ref{fig0126.9} a, b and d).
This ionized region expands with the growth of the protostar 
owing to the increasing radiative heating.
Next, we see the case of $980M_{\odot}$, where the flow is remarkably decelerated by the radiation force.
Already at the outer boundary, the inflow is slow with only 20\% of the free-fall velocity.
Subsequently, the flow is decelerated to 10\% of the free-fall velocity
by the radiation force at $5 \times 10^{3} \rm{au}$.
Since the envelope is very optically thin, the radiation force is dominated by the direct radiation.
Further inside, although the flow is once accelerated after the dust sublimation
and reaches almost the free fall velocity around $100 \rm{au}$
as a result of the decrease of opacity and so radiation pressure,
the ionization of gas raises the opacity near the star ($<10  \rm{au}$)
and the flow becomes as slow as 40\% of the free fall again.
Note also the density enhancement accompanying the slow down of the flow 
in the outer ($\ga 5 \times 10^{3} \rm{au}$) and 
inner ($\la 10 \rm{au}$) regions (Figure \ref{fig0126.9} c).

Figure \ref{fig0126.11} shows the local optical depths for the direct and diffuse light.
Again those values decrease with the protostellar mass 
except for the final epoch of $M_* \simeq 980~M_{\odot}$.
The envelope is optically thin to the diffuse light for any mass range
and also to the direct light for $M_* \ga 15M_{\odot}$.

Figure \ref{fig0126.10} shows the radiative energy distribution at different radii
for the final epoch $M_* \simeq 980M_{\odot}$. 
The dust destruction front locates at $\sim 10^{3} \rm{au}$
and the inner three radii shown ($1,10,10^2,10^3\rm{au}$) correspond to the dust-free region
and the outermost one ($10^{4}\rm{au}$) is in the dusty region.
In the inner dust-free region, the radiation field is dominated by 
the direct light in the black-body shape
peaking at $0.3 \mu \rm{m}$ ($10^{15} \rm{Hz}$).
From the comparison between the spectra in the dust-free ($10 \rm{au}$)
and in the dusty ($10^{4} \rm{au}$) regions,
we see that the direct light is hardly absorbed in the envelope although with slight
modification.
For example, the bump around $10 \mu{\rm m}$ in the spectrum at $10^{4} \rm{au}$
is due to the re-emitted radiation by the dust grains as in Figure \ref{fig0126.6}.

\subsection{Comparison between numerical and analytical results}\label{kekka_hikaku}

In our spherically symmetric steady-flow framework, we
cannot follow the evolution until the flow completely halts. 
Here, as a guide, we choose the epoch when the flow velocity falls to 1/10 of the free-fall
velocity as the end of the accretion and obtain the stellar upper mass
limit. 
The upper mass limit obtained in this way has been presented in Figure~\ref{fig0127.1}.
Somewhat different choice of this criterion changes the result
only moderately since the protostellar luminosity and thus its
radiation pressure on to the flow increases rapidly beyond this mass.
If we continued calculation after this point, 
the flow would be ever decelerated and more gas accumulates 
in a shell-like structure.
In such a situation, the thermal pressure, which is neglected here, 
can be important as the flow becomes subsonic (see Figures  \ref{fig0126.1} and \ref{fig0126.5}). 
Although we can imagine the flow solutions where
thermal pressure force in the shell pushes the gas 
inward against the radiation force, 
this is an artifact of spherical symmetry assumption. 
The shell-like structure would not be maintained 
in realistic multi-dimensional situation by growth of non-spherical perturbation 
 \citep[e.g.,][]{Krumholz2009} . 
Here, since our aim is to estimate the epoch when the stellar radiation feedback on to 
the flow becomes remarkable and we are not interested in 
detailed shell structure when the flow is very decelerated,
our neglect of thermal pressure is justified.

The different symbols in Figure~\ref{fig0127.1} represent difference 
in the way the radiation force works on
the flow in the final stage, i.e.,
(i) the dust cocoon is optically thick also to the diffuse light
and deceleration of the flow is mainly caused by diffuse light,
(ii) the dust cocoon is optically thick to the direct light but thin to the diffuse light,
and deceleration is mainly by the direct light at the dust destruction front,
and (iii) the dust cocoon is optically thin also to the direct light
and deceleration by direct light occurs in the entire dust cocoon.
In addition, at very low metallicities where the HII region expansion
is predicted by the analytic argument
to be more effective than the radiation force in halting the accretion,
the analytic estimate by Equation  (\ref{0330.2.1}) is plotted instead as the upper limit.
Here, when the cocoon is optically thick to the direct light,
we have classified the cases (i) or (ii) according to
whether the contribution to the radiation force is dominated
by the direct light or by the diffuse light
at the radius $r_{\rm rad}$ where the ratio of radiation pressure and gravity becomes maximum.
Since the former is $\propto \kappa_{\rm UV} e^{-\tau_{\rm rad}}$, where,
$\tau_{\rm rad}$ is the optical depth between the dust destruction front and $r_{\rm rad}$ for direct light,
and the latter is $\propto \kappa_{\rm IR} (1 - e^{ - \tau_{\rm rad}})$, the case (i), i.e.,
the former is higher than the latter, corresponds to
$\tau_{\rm rad}  >  \ln ({\kappa_{\rm UV}/\kappa_{\rm IR} + 1}) \simeq 4.3$, and vice versa.
Here, we have used $\kappa_{\rm UV}/\kappa_{\rm IR} = 70$ from Section \ref{dust_press}.

\begin{figure}
	\begin{center}
		\includegraphics[width=\columnwidth]{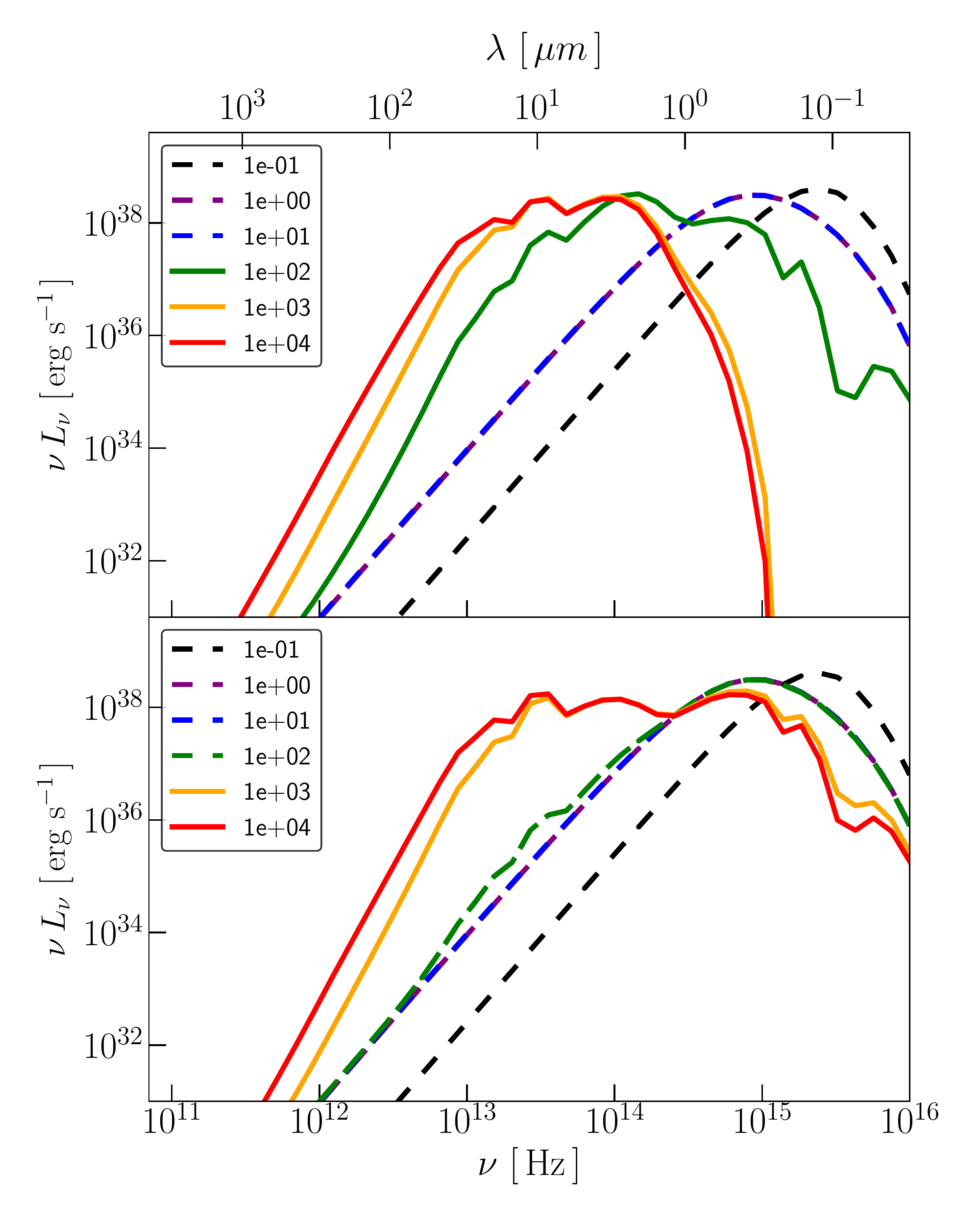}
	\end{center}
	\caption{Comparison of the spectral energy distribution at $57M_{\odot}$
with metallicities $Z = 10^{-1}Z_{\odot}$ (top) and $10^{-2}Z_{\odot}$ (bottom).   
In both cases, the cocoon is optically thick to the UV ($\tau_{\rm UV} > 1$) 
but thin to the IR wavelengths ($\tau_{\rm IR} < 1$). 
The lines are for those at the radii, $0.1 - 10^{4} \rm{au}$, indicated in the legend.
The solid and dashed lines correspond to the region where the dust is present and absent, respectively.}
	\label{fig0127.2}
\end{figure}

The numerical and analytical results shown in Figure \ref{fig0127.1}
are in good agreement each other in the cases that
(i) the dust cocoon is optically thick for diffuse light
and (iii) optically thin for direct light,
but there is some discrepancy
in the case (ii) where the flow is stopped by radiation pressure of direct light
at the dust destruction front.
In particular,
in the case (ii), the analytic estimate for the upper mass limit does not depend on the
metallicity (Eq. \ref{0122.2}), while
the numerical ones increase toward lower metallicities.
This comes from breakdown of the assumption in the analytical estimate
that all the direct light is absorbed at the dust destruction front (Eq. \ref{0122.1}).
In reality, a part of the radiation escapes without being absorbed
in low-metallicity cases, where the optical depth is relatively small.
Toward the lower metallicity, the momentum received by the accretion flow
from the direct light thus decreases so that the upper mass limits increases. 
To illustrate this effect, we show the radiative energy distributions
in the two cases where $\dot M = 10^{-3}M_{\odot} \rm{yr^{-1}}$, $M_{*} = 57M_{\odot}$, $Z = 10^{-1}Z_{\odot}$ and $10^{-2}Z_{\odot}$ in Figure \ref{fig0127.2}.
In both cases, the dominant radiative feedback is by the radiation force at the dust destruction front.
For $10^{-1}Z_{\odot}$, we can see that the direct light peaking at $0.3\mu \rm{m}$ 
($10^{15}\rm{Hz}$)
is totally absorbed and re-emitted in the dust cocoon, and the radiation field
outside $10^{3}\rm{au}$ consists only of the diffuse component (top panel of Figure \ref{fig0127.2}).
On the other hand, in the case of $10^{-2}Z_{\odot}$, as seen in the spectra at
$10^{3}$ and $10^{4} \rm{au}$, a part of the direct light escapes from the dust cocoon
without absorption (bottom panel of Figure \ref{fig0127.2} ).
The dust opacity has frequency-dependence as shown in Figure \ref{fig1.1} and decreases toward longer wavelengths from $0.3 \mu {\rm m}$ where energy distribution of the direct light becomes maximum, e.g., the absorption opacity  becomes $\kappa \approx 70{\rm cm^{2}{g^{-1}}}  \left( Z/Z_{\odot} \right)$ at $1 \mu {\rm m}$.
This value is 1/5 smaller than the opacity, $\kappa_{\rm UV} = 350{\rm cm^{2}{g^{-1}}} \left( Z/Z_{\odot} \right)$, used for the analytical estimate for the optical depth, $\tau_{\rm UV}$, in Section \ref{sec.kaiseki}.
Therefore, the cocoon becomes optically thin for the direct light at $\sim 1\mu {\rm m}$ even if the optical depth is $\tau_{\rm UV} \approx1$ in our analytical framework.
In this case, radiation pressure becomes weaker than estimated in Equation \eqref{0122.1} because longer-wavelength photons can escape through the cocoon.
In the analytical estimate, Equation \eqref{0122.1}  is only applicable when the dust cocoon becomes optically thick for all main components of the direct light.   
For the cocoon to be optically thick also to those long-wavelength photons 
around $1\mu{\rm m}$ and Equation \eqref{0122.1} to be usable,  
$\tau_{\rm UV}>5$  is required in the analytical estimate.

\section{Results for cases with metallicity-dependent 
accretion histories}
\label{sec.tenkei}

So far, the accretion rate is treated as a free parameter and arbitrarily chosen 
for a given metallicity.
In reality, the accretion rate should depend on the temperature during the pre-stellar collapse
(Eq. \ref{1117.1}), which is controlled by the amount of dust grains
and heavy elements, i.e., metallicity.

\begin{figure}
        \begin{center}
                \includegraphics[width=\columnwidth]{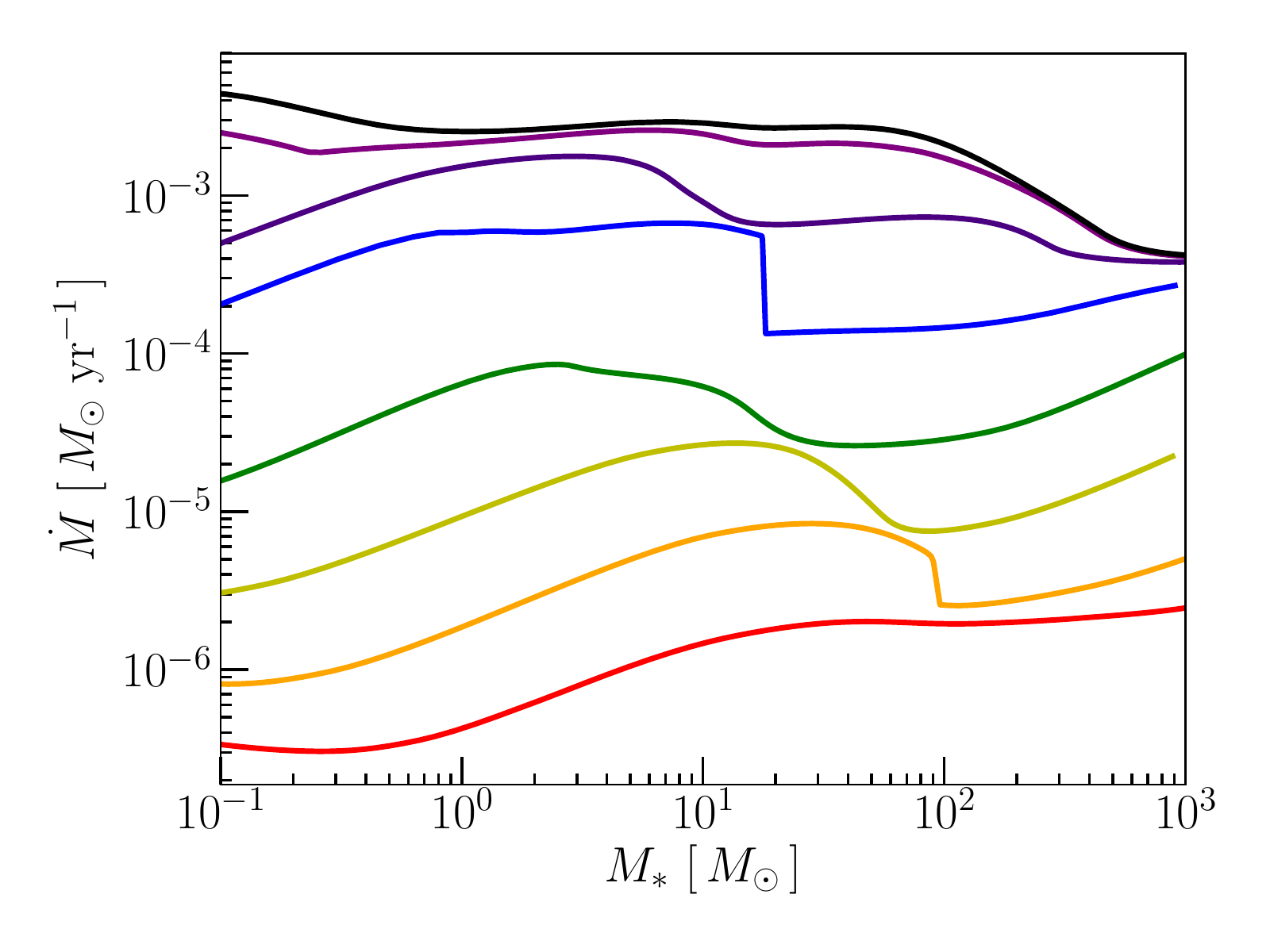}
        \end{center}
        \caption{ The accretion-rate histories for metallicities $1$,$10^{-1}$,$10^{-2}$,$10^{-3}$,$10^{-4}$,$10^{-5}$,$10^{-6}$, $0 Z_{\odot}$ (from bottom to top) 
derived from pre-stellar temperature evolution. Adapted from HO09b.}
        \label{fig0128.1}
\end{figure}

\begin{figure}
        \begin{center}
                \includegraphics[width=\columnwidth]{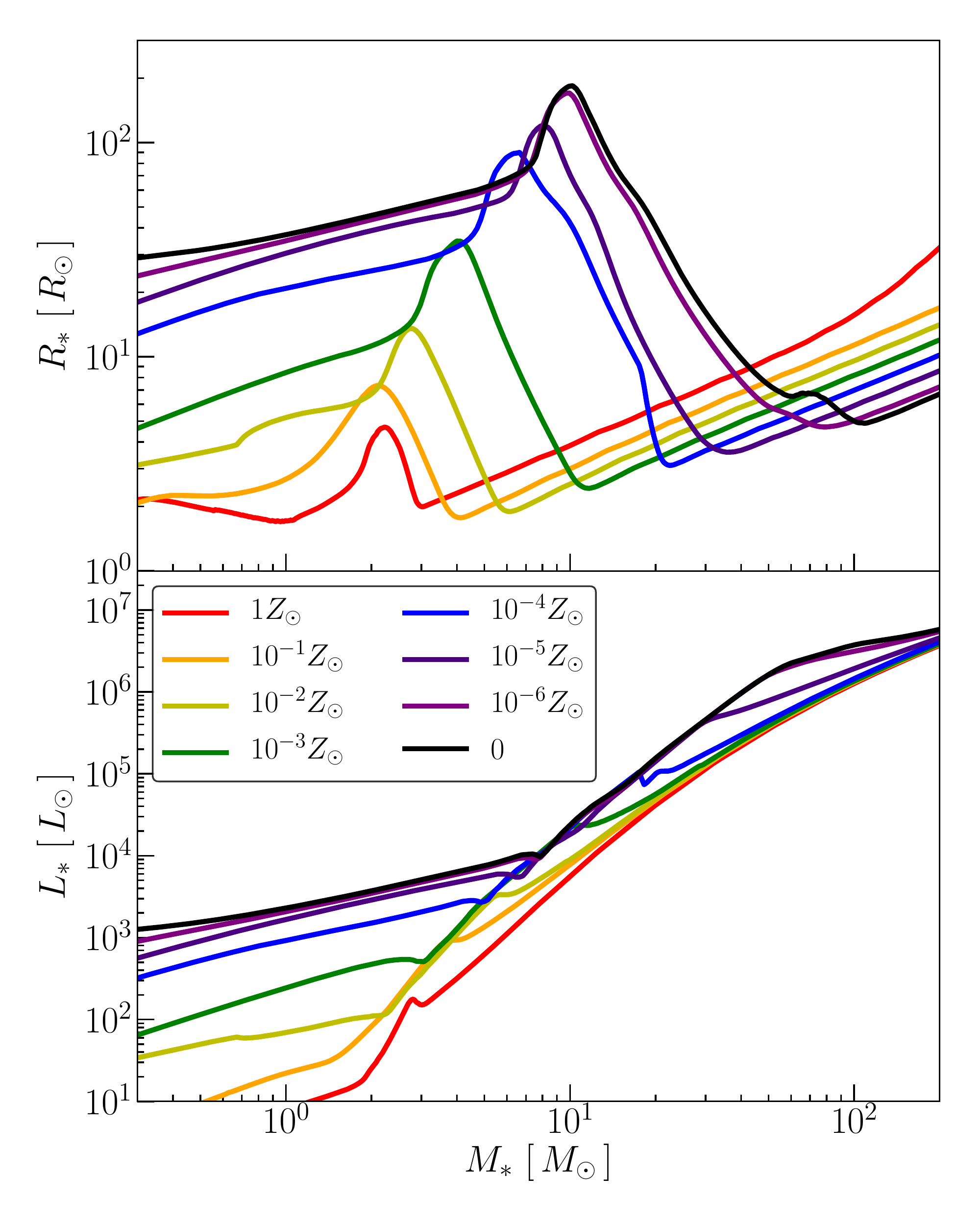}
        \end{center}
        \caption{The evolution of protostellar radii (top) and luminosities (bottom) 
for different metallicities, as shown in Figure \ref{fig0119.1}, but for accretion histories given in 
Figure \ref{fig0128.1}.}
        \label{fig0129.1}
\end{figure}

In this section, we use accretion history derived from temperature evolution during the collapse
for each metallicity and estimate the upper mass limit as a function of metallicity.
The accretion rates adopted here are shown in Figure \ref{fig0128.1}, which is
the same as in HO09b, who
derived those rates from the temperature evolution by the one-zone 
calculation of \citet{Omukai2005}.
With those accretion histories, we solve protostellar evolution 
as in Section \ref{stellar.evlol}. The obtained evolution of the
stellar radii and luminosities are summarized in Figure \ref{fig0129.1}.

\begin{figure}
        \begin{center}
                \includegraphics[width=\columnwidth]{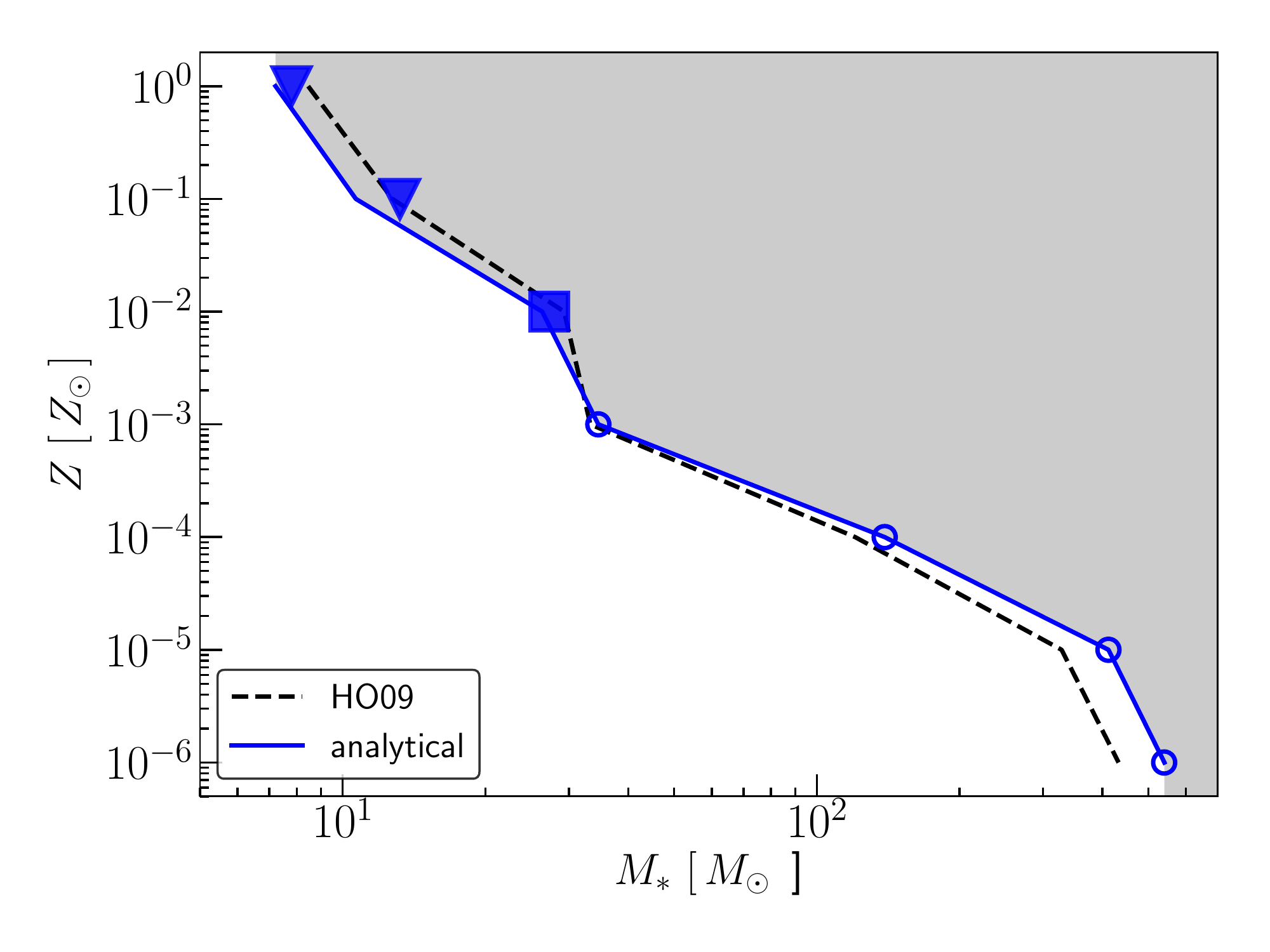}
        \end{center}
        \caption{The stellar upper mass limit at different metallicities 
for the accretion histories shown in Figure \ref{fig0128.1} and the protostellar evolution 
in Figure \ref{fig0129.1}. 
Symbols represent the limit estimated from the numerical models. 
The symbols indicate the dominant mechanisms for halting the inflow at the limits 
as in Figure \ref{fig0127.1}:
(1) the cocoon is optically thin to the diffuse radiation but still thick to the direct radiation, 
and the flow is decelerated mostly by the direct radiation pressure around the dust destruction front
($\triangledown$);
(2) the cocoon is optically thin also to diffuse radiation, and so the direct radiation pressure onto the entire cocoon 
decelerates the flow ($\Box$).
Also shown by the blue line is the limit by the analytical argument in Section \ref{sec.kaiseki}.
For comparison, the analytic estimate by HO09b is shown by the dashed line.
Note that, in the case that the maximum mass is limited by the HII-region growth, 
the symbols $\bigcirc$ indicate the analytical estimate given in Section \ref{HII_region}}
\label{fig0129.2}
\end{figure}

The upper mass limit by the analytical argument in Section \ref{analytical_upper_mass}
is presented in Figure \ref{fig0129.2} for the accretion histories shown in Figure \ref{fig0128.1}.
The meaning of lines are the same as in Figure \ref{fig0124.1},
but the limits are calculated only at metallicities
where the accretion histories are given.
In Section \ref{sec.kaiseki} (Figure \ref{fig0124.1}), at relatively high metallicity,
the diffuse radiation force is important in terminating accretion.
Here, however, such a situation does not appear for any metallicity
since the accretion rate is very small $<10^{-6}M_{\odot}\rm{yr^{-1}}$
in cases with relatively high metallicity ($\geq 10^{-1}Z_{\odot}$)
and the envelope is optically thin to the diffuse light.
In such cases, the dust cocoon is optically thick to the direct light
and the radiation force at the dust destruction front sets
the upper mass limit at $\sim 10M_{\odot}$ by Equation \eqref{0122.2}.
With metallicity as low as $10^{-2}Z_{\odot}$, the dust cocoon becomes 
transparent also to the direct light.
Thus, the direct light exerts the radiation force in the entire dust cocoon,
setting the upper mass limit at $26M_{\odot}$ by Equation \eqref{0123.2}.
With even lower metallicity of $\leq 10^{-3}Z_{\odot}$,
the dominant feedback is by the expansion of the HII region,
and the upper limit is given by Equation \eqref{0330.2.1},
which depends only on the accretion rate.
With lower metallicity,
the accretion rate becomes higher and the upper mass limit by the HII-region expansion increases, 
which is $34M_{\odot}$ at $10^{-3}Z_{\odot}$, and $540M_{\odot}$ at $10^{-6}Z_{\odot}$, respectively.
In Figure \ref{fig0129.2}, the analytic estimate by HO09 is also shown (dashed line).
This is very close to our numerical estimate, but there is slight differences.
Firstly, HO09 set the dust destruction front radius
at $35 \left( L_{*}/ 10^{5}L_{\odot} \right)^{1/2} \rm{au}$, 
which is smaller than the value used here. This results in slightly
lower upper limit than ours (Eq. \ref{0122.2}) at $Z_{\odot}$.
Secondly, HO09 assumed the free-fall density distribution
in estimating the effect of the HII region expansion, while we here take into account
the radiation pressure effect on to the flow and so the density is enhanced from 
theirs (Eq. \ref{eq4.8}).
Thus, the upper mass limit becomes higher in our case because more ionizing photons are
needed for the expansion of HII region.
However, both of the above have just minor effects.

As in Section \ref{sec.kekka_env}, we calculate the envelope 
evolution numerically to find the upper mass limits, 
which are shown in Figure \ref{fig0129.2} by the same symbols as in Figure \ref{fig0127.1}.
With relatively high metallicities of $Z = 1~Z_\odot$ and $10^{-1}Z_{\odot}$,
the upper limits are $7.8$ and $13M_{\odot}$ respectively,
set by the radiation pressure at the dust destruction front.
At metallicity of $10^{-2}Z_{\odot}$,
the dust cocoon becomes transparent also to the direct light,
which finally limits the accretion at $M_* \simeq 27M_{\odot}$
via the radiation pressure working on the entire dust cocoon.
At metallicity $10^{-3}Z_{\odot}$, the radiation feedback
does not become significant before reaching $35M_{\odot}$,
the analytical estimate for the maximum mass by the HII region expansion.
With even lower metallicities, the upper limits are all set by the HII region feedback.
Note that the upper limits by the radiative feedback agree surprisingly well
with the analytical estimates in all the metallicity range.
At metallicity $10^{-1}Z_{\odot}$,
although the optical depth to the direct light is close to unity, and so
the numerical result is larger than the analytical value
as mentioned in Section \ref{kekka_hikaku}, this difference remains very small.


\section{Conclusion and Discussion}\label{sec.diss}

We have calculated the evolution of accreting protostars
and its envelope structures for different metallicities
assuming the spherically symmetric steady accretion.
By regarding the moment when
the strong radiative feedback decelerates 
the accretion flow down to 10\% of the free-fall velocity 
as the epoch when the accretion ceases,
we have derived the upper stellar mass limits.
First we have investigated the cases with various parameter sets of
accretion rates $\dot M=10^{-3}$,$10^{-4}$,$10^{-5}M_{\odot} \rm{yr^{-1}}$ 
and metallicities $Z=1$, $10^{-1}$, $10^{-2}$, $10^{-3}$, $10^{-4}Z_{\odot}$.
We then have studied the cases where the accretion histories are determined 
by the thermal evolution in the prestellar collapse, which varies 
with different metallicities. 
We have also compared the numerical results to our previous analytic estimates
, and found that the analytical estimates are in good agreement with 
the numerical results.
Our findings can be summarized as follows.

\begin{itemize}
\item[--] The upper mass limit is set by the radiation force on to the dust cocoon
in relatively metal-enriched cases with $\ga 10^{-3}-10^{-2}Z_{\sun}$,
and increases toward lower metallicity for a fixed constant accretion rate.
At even lower metallicities, the HII region expansion terminates the mass accretion
and the upper limit in this case does not depend on the metallicity for the
same accretion rate.

\item[--] At the constant accretion rate of $10^{-4}M_{\odot}\rm{yr^{-1}}$, which is roughly the
typical value for the present-day high-mass star formation
as well as in primordial star formation,
the radiation force on to the dust cocoon terminates the protostellar accretion when the
protostar becomes massive enough in cases with metallicity higher than $\sim 10^{-2}Z_{\odot}$.
The upper mass limit increases with decreasing metallicity
from $12M_{\odot}$ (at $1Z_{\odot}$), $30M_{\odot}$ (at $10^{-1}Z_{\odot}$) to 
$47M_{\odot}$ (at $10^{-2}Z_{\odot}$).
The radiative feedback can be classified in to three cases according to the optical thickness
of the dusty envelope.
With high enough metallicity ($\sim 1Z_{\odot}$), the dusty envelope is optically thick
not only to the direct stellar light but also to the re-emitted infrared light.
In such cases, the radiation feedback is dominated by the re-emitted light and
its effect on the flow is most important outside the dust destruction front.
With lower metallicity ($\sim 10^{-2}- 0.1 Z_{\odot}$),
the dusty envelope becomes optically thick to the direct light but is still thin to the diffuse light.
In this case, the radiation force by direct light at the dust destruction front works as the dominant
feedback mechanism to terminate the accretion.
With metallicity below $\sim 10^{-2} Z_{\odot}$, the envelope becomes optically thin also to the diffuse
light and the direct radiation force is exerted to the entire dusty envelope.
At such low metallicities, the expansion of the HII region is actually  
more effective in terminating the accretion.
The upper mass limit by this mechanism is $94M_{\odot}$ for this accretion rate
and does not depend on the metallicity.

\item[--] In general, the upper limit increases with the accretion rate 
for a fixed metallicity (see Figure \ref{fig0127.1}).
At metallicity $\sim 1Z_{\odot}$, where the dominant feedback is due to the diffuse light, 
the maximum protostellar mass is, however,
always around $20M_{\odot}$ and does not depend on the accretion rate (see Eq. \ref{0123.1} ).

\item[--] When the accretion-rate history estimated
from the prestellar thermal evolution at each metallicity is used, 
the maximum mass increases more rapidly toward lower metallicities
than in the case of constant accretion rate
since the accretion rate also becomes higher at lower metallicities.
In this case, the radiation feedback to the dust cocoon sets the maximum mass
for $\ga 10^{-2}Z_{\odot} $ and this limit increases from $8M_{\odot}$ (at $1Z_{\odot}$)
to $30M_{\odot}$ (at $10^{-2}Z_{\odot}$).
Below $\sim 10^{-2}Z_{\odot}$, the HII-region expansion instead limits 
the stellar mass growth
and the maximum mass increases from
$35M_{\odot}$ at $10^{-3}Z_{\odot}$ to $540M_{\odot}$ at $Z=0$
as a result of the increase in the accretion rate.
Our numerical results in this case are in very good agreement 
with the analytic estimate by HO09.
\end{itemize}

\hspace{8pt}

Although with simplistic assumptions of the spherically symmetric 
and steady state accretion, the upper mass limits obtained here would 
give a rough guide for the epoch when accretion growth of stars
is strongly hindered by the stellar feedback.
Based on this result, we here discuss possible variations 
of the stellar mass as a function of metallicity.
At metallicity as high as $\sim 1Z_{\odot}$, regardless of the accretion rate,
stars more massive than $20M_{\odot}$ cannot be formed by the spherical accretion.
In contrast, for metallicity $Z \la 10^{-1}Z_{\odot}$, stars more massive than dozens of the sun can easily be formed
as long as the accretion rate is higher than $10^{-4}M_{\odot} \rm{yr^{-1}}$.
Somewhat dramatic jump in the maximum mass occurs at metallicity between $10^{-2}Z_{\odot}$ and $10^{-3}Z_{\odot}$:
it is $\simeq 100M_{\odot}$ in the range of $10^{-2}-10^{-1}Z_{\odot}$, 
but reaches $\simeq 10^{3}M_{\odot}$ at $\la 10^{-3}Z_{\odot}$
(see Figures \ref{fig0124.1} and \ref{fig0127.1}).
This jump in the upper mass limit might also be imprinted in the shape of the stellar IMF at this metallicity range.
Observationally, the slope of the IMF of star clusters and OB associations in low-metallicity environments
such as in the Magellanic Clouds is very similar to that in the Milky Way
 \citep[e.g.,][]{Bastian2010}.
The metallicity effects on IMF might just be masked or compensated by other factors affecting the stellar mass as well
such as the fragmentation properties during the prestallar collapse, turbulence  etc. \citep{Kroupa}.
Also, while no star with the initial mass higher than $200M_{\odot}$ has not been found in our Galaxy,
those with more than $300M_{\odot}$ appear to exist in the Large Magellanic Cloud \citep{Crowther_2010}.
This difference in the cut-off in the IMF may come from the metallicity effect on the stellar feedback.

With even lower metallicity of $\la 10^{-3}Z_{\odot}$, the maximum mass is limited by
the HII-region expansion and depends only on the accretion rate, rather than on the metallicity.
In this case, the accretion evolution of protostars becomes almost identical 
to that of the primordial stars, for which three dimensional simulations 
predict that very massive stars with $M_* \gtrsim 600M_{\odot}$ can be formed
\citep{Hosokawa2016}.
Nonetheless, the effects of finite metallicity may
change the evolution. For instance, presence of a small amount of
heavy elements will modify the structure of the
HII region, via additional cooling and absorption of UV photons.
We will take these effects into account in future studies.


\begin{figure}
        \begin{center}
                \includegraphics[width=\columnwidth]{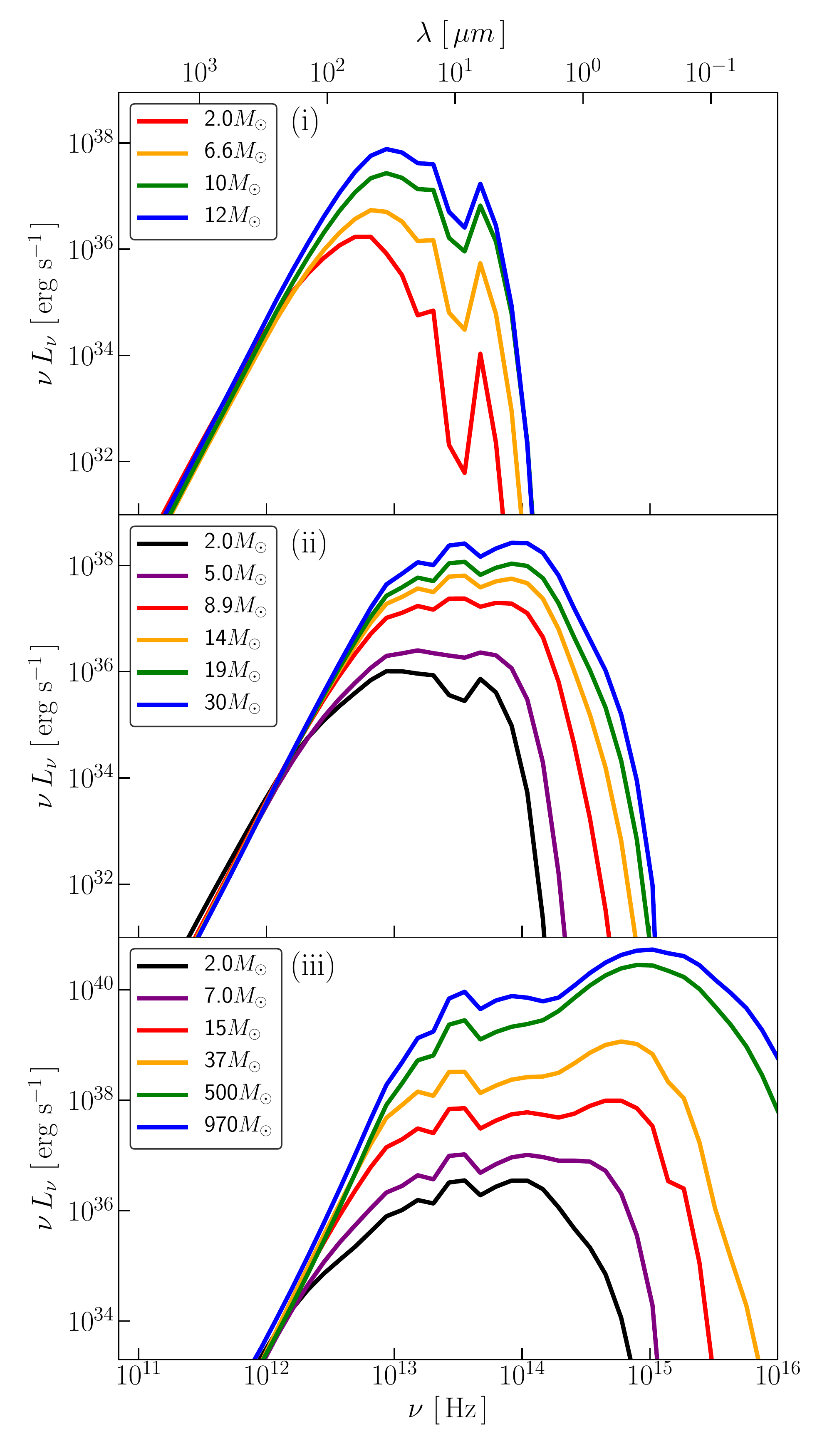}
        \end{center}
        \caption{Evolution of the spectral energy distribution observed from the outside. 
The lines in each panel correspond to different epochs. 
Panel (i) shows the case of $Z=1Z_{\odot}$ and $\dot M = 10^{-4}M_{\odot} \rm{yr^{-1}}$, 
where $\tau_{\rm UV},\tau_{\rm IR} > 1$, as seen in Section \ref{IR_thick}, 
(ii) the case of $Z = 10^{-1}Z_{\odot}$ and $\dot M = 10^{-4}M_{\odot}\rm{yr^{-1}}$, where
$\tau_{\rm UV} > 1,\tau_{\rm IR} < 1$, as seen in Section \ref{dest_ruct}, 
and (iii) the case of $Z = 10^{-3}Z_{\odot}$ and $\dot M = 10^{-3}M_{\odot}\rm{yr^{-1}}$, where
$\tau_{\rm UV},\tau_{\rm IR} < 1$, as seen Section \ref{opticall_thin_for_UV}. }
        \label{fig0424}
\end{figure}

The way that the radiation force disturbs the accretion flow qualitatively
differs with different optical depths of the dust cocoon, 
as seen in this paper. 
We can investigate whether the stellar direct light is
converted to the diffuse light through the dust cocoon to
terminate the accretion by observing the spectral energy 
distributions of protostars. 
Figure \ref{fig0424} shows such spectral features
of accreting protostars seen from outside
for three optical-depth regimes: (i) $\tau_{\rm UV},\tau_{\rm IR} > 1$,
(ii) $\tau_{\rm UV} > 1,\tau_{\rm IR} < 1$, and (iii) $\tau_{\rm UV},\tau_{\rm IR} < 1$.
In the optically thick case shown in the panel (i),
the temperature of the dust photosphere is low at a few 100 K
and the spectral peak is around $30\mu \rm{m}$ through the protostar phase.
The $10\mu \rm{m}$ silicate feature by the stretching mode of the Si-O bond appears as a dip
in the spectrum.
In the intermediate optical depth case shown in panel (ii), the radiation is emitted from the dust destruction front
at $1200\rm{K}$ and the spectrum peaks at $\simeq 4 \mu \rm{m}$.
In the mid-infrared, the spectrum is very flat with the $10\mu \rm{m}$ feature appearing as a bump, unlike in panel (i).
In the optically thin case shown in panel (iii), the direct light from the protostar dominates
the spectrum, peaking at the UV to optical bands. The $10\mu \rm{m}$ feature appears as a bump
as in (ii).

Finally, we discuss a few effects which were not included in our numerical modeling.
Although we have regarded the moment that the flow velocity in the envelope falls below
1/10 of the free-fall value at some radius as the termination epoch of the accretion,
strictly speaking the stars can continue to grow even after that 
in our spherical accretion framework.
From the fact that the luminosity increases rapidly and radiation feedback becomes
correspondingly intense toward higher masses, we expect that
the accretion rate is greatly reduced slightly after the stellar mass
exceeds our upper limit and the accretion will be stopped before long.
Nonetheless, multi-dimensional effects must be important. In fact,
stars much larger than our spherical limits are found even in the solar neighborhood.
This discrepancy is just the radiation-pressure barrier described in Section \ref{introduction}, 
and will be attributable to the three-dimentional nature of the accretion.
In a realistic situation, a circumstellar disk is formed due to the
finite angular momentum and accretion proceeds through the disk.
Radiation thus escapes preferentially in the polar directions perpendicular to the disk,
and the radiative feedback to the matter behind the disk becomes weaker.
In addition,
even the gas in the polar direction, which receives more direct light, might be able to accrete
owing to the Rayleigh-Taylor instability at the dust destruction front
\citep[but see \citealt{Klassen2016,Kuiper2012}]{Rosen2016,Krumholz2009}.
As a result of these processes, stars much more massive than the spherical limit could be formed
\citep{Yorke_Bodenheimer1999,Krumholz2009,Kuiper2012,Tanaka2017}.
Those multi-dimensional studies are mostly for $\sim 1Z_{\odot}$, 
and further studies are still awaited for the cases with the lower metallicities.

We have assumed that dust grains are dynamically coupled with the gas.
The radiation force on dust grains is transferred to the gas via drag force by collision.
The assumption of the gas and dust coupling breaks down
if the velocity offset between the dust and the gas becomes comparable to the flow velocity.
We here estimate their velocity offset $v_{\rm d}$ considering
the balance between the radiation and drag forces on the grains \citep{Wolfire_Cassinelli1987}.
Note that the gravity on grains is negligible compared to the radiation force.
Radiation force onto a grain with radius $a$ is
\begin{eqnarray}
F_{\rm rad} = \frac{\pi a^2 \int L_{\nu} Q_{\nu} d \nu}{4 \pi r^2 c} = \frac{\pi a^2 L_{*} \langle Q \rangle}{4 \pi r^2 c},
\label{0421.1}
\end{eqnarray}
where $Q_{\nu}$ is the extinction efficiency factor and $\langle Q \rangle =  \int L_{\nu} Q_{\nu} d \nu / L_{*}$.
Drag force $F_{\rm drag}$ is given by \citep{Draine_Salpeter},
\begin{eqnarray}
F_{\rm drag} = 2 \pi a^2 k T_{\rm g} \frac{\rho}{\mu m_{\rm H}} \frac{8}{3 \sqrt{\pi}} S \left( 1 + \frac{9 \pi}{64} S^2 \right)^{1/2},
\label{0421.3}
\end{eqnarray}
where $S = \left(  \frac{\mu m_{\rm H}}{2 k T_{\rm g}} v_{\rm d}^2 \right)^{1/2} $.
When the grains completely decouple from the gas, the relative velocity becomes supersonic 
and so $S \gg 1$.
From $F_{\rm rad}=F_{\rm drag}$, the relative velocity $v_{\rm d}$ becomes
\begin{eqnarray}
	v_{\rm d} = \sqrt{\frac{L \langle Q \rangle}{4 \pi r^2 c \rho}}. \label{0421.5}
\end{eqnarray}
or in terms of the ratio with the free-fall velocity (Eq. \ref{0331.1}),
\begin{align}
	v_{\rm d} / u_{\rm ff} &= 1.26  \left( \frac{L_{*}}{8 \times 10^5 L_{\odot}} \right)^{1/2} \left( \frac{\langle Q \rangle}{1} \right)^{1/2} \left( \frac{\dot M}{10^{-3}M_{\odot} \rm{yr^{-1}}} \right)^{-1/2} \nonumber \\
		& \left( \frac{M_{*}}{60M_{\odot}} \right)^{1/4} \left( \frac{R}{10^3 \rm{AU}} \right)^{-1/4} .  \label{0421.7}
\end{align}
The value of $\langle Q \rangle$ differs greatly between the direct and the diffuse light.
For example, graphite grains of $0.1\mu \rm{m}$ have $\langle Q \rangle \approx 1$
for the direct light at $2 \times 10^4 \rm{K}$, while $\langle Q \rangle \approx 10^{-2}$ for the diffuse light at $500  \rm{K}$.
This means that the decoupling between the dust and gas occurs when the dust envelope is optically thin for the direct light.
In this case, radiation force on to the flow is reduced from our estimate
because the grains are removed from the envelope and the opacity decreases.
Without radiation feedback onto the dust cocoon,
the maximum mass in this case would be set by the HII-region expansion.
Recall, however, that optically-thin direct light terminates the accretion only
in a very limited range in metallicity (see Figs. \ref{fig0127.1} and \ref{fig0129.2}).
Note also that the drift velocity by Equation \eqref{0421.7} is an upper limit
since the Coulomb drag effect \citep{Draine_Salpeter} is not considered.
For charged grains, larger drag by this effect keeps the gas-dust coupling well.
For those reasons, we expect that the gas-dust decoupling does not affect 
the upper mass limit in most of the metallicity range.

\section*{Acknowledgements}
The authors wish to express our cordial thanks to Prof. Takahiro Tanaka 
for his continual interest and advice.
We also thank Daisuke Nakauchi, Kazu Sugimura, Sanemichi Takahashi, 
Hidekazu Tanaka and Hide Yajima for fruitful discussions.
This work is supported in part by MEXT/JSPS KAKENHI 
Grant Number (KO:25287040, 17H01102, TH:16H05996).


 \begin{table*}
  \caption{Fitting Coefficients For $\alpha$ ( Eq. \ref{ap2.3})}
  \label{a1tab}
  \begin{tabular}{ccccccc}
    \hline \hline
     Temperature ( K ) &  $a_{0}$ & $a_{1}$ & $a_{2}$ & $a_{3}$ & $a_{4}$ & $a_{5}$  \\ 
    \hline
    T < 1000 & $9.78382 \times 10^{-1}$ & $3.99572 \times 10^{-4}$ & $-1.73108 \times 10^{-6}$ & $1.15363 \times 10^{-9}$ & $8.24607 \times 10^{-13}$ & $-7.65975 \times 10^{-16} $\\
    1000 < T < 4000 & $8.27472 \times 10^{-1}$ & $1.07697 \times 10^{-4} $ & $ -8.25123 \times 10^{-8} $ & $1.92812 \times 10^{-12}$ &  $5.7192 \times 10^{-15}$ & $-7.869 \times 10^{-19} $ \\
    4000 < T  & $1.11569$ &  $- 3.29302 \times 10^{-4} $ & $1.01846 \times 10^{-7}$ & $-1.46666 \times 10^{-11} $  & $1.00764 \times 10^{-15} $ & $ -2.68873 \times 10^{-20} $ \\
    
    \hline
  \end{tabular}
 \end{table*}

 \begin{table*}
  \caption{Fitting Coefficients For $N_{\rm c}$ ( Eq. \ref{ap2.4})}
  \label{b1tab}
  \begin{tabular}{cccccc}
    \hline \hline
      $b_{0}$ & $b_{1}$ & $b_{2}$ & $b_{3}$ & $b_{4}$ & $b_{5}$ \\ 
    \hline
      $24.0561$ &  $1.10043 \times 10^{-3}$ & $-2.87224 \times 10^{-7}$ & $6.11525 \times 10^{-11}$ & $-6.55034 \times 10^{-15}$  & $2.54997 \times 10^{-19}$  \\
    \hline
  \end{tabular}
 \end{table*}






\appendix
\section{Method for radiative transfer calculation}\label{apdi1}

The source function $S_{\nu}$ in the momentum equation \eqref{1118.1} is given by Equation \eqref{1118.4}, which contains $B_{\nu}(T)$. 
To accelerate the convergence of iterative calculation, we need to give the source function as an explicit function of $J_{\nu}$ 
in the following way \citep{Mihalas_Mihalas1984}.
By rewriting Equation \eqref{eq1.6} as
\begin{eqnarray}
		\int \kappa_{\nu}^{\rm abs} B_{\nu} d \nu = \int \kappa_{\nu}^{\rm abs} J_{\nu} d \nu  - u \frac{de}{dr} - P u \frac{d}{dr} \left( \frac{1}{\rho} \right)  + \Lambda_{\rm chem} + \Lambda_{\rm line}   \label{1125.1}
\end{eqnarray}
and using Equation \eqref{1125.1}, we can write the source function (Eq. \ref{1118.4}) as 
\begin{align}
	S_{\nu} &= \frac{1}{\kappa_{\nu}^{\rm abs} + \sigma_{\nu}^{\rm sc}}  \frac{\kappa_{\nu}^{\rm abs} B_{\nu}}{\int \kappa_{\nu}^{\rm abs} B_{\nu} d \nu} \nonumber \\
	 & \times \biggl[ \int \kappa_{\nu}^{\rm abs} J_{\nu} d \nu - u \frac{de}{dr} - P u \frac{d}{dr} \left( \frac{1}{\rho} \right) + \Lambda_{\rm chem} + \Lambda_{\rm line} \biggr] \nonumber  \\ 
& + \frac{\sigma_{\nu}^{\rm sc} J_{\nu}}{\kappa_{\nu}^{\rm abs} + \sigma_{\nu}^{\rm sc}}.  \label{1125.2}
\end{align}
With Equation  \eqref{1125.2}, the source function $S_{\nu}$ is evaluated to satisfy Equation \eqref{1125.1} about $J_{\nu}$  when we integrate the moment equations of the radiation transfer equations (Eq. \ref{1118.1} and \ref{1118.2}). 
 In this method, the effect of changing the value of $J_{\nu}$ is instantly reflected to the source function $S_{\nu}$, so the convergence of solution becomes faster.

We calculate the envelope structure in the following way. 
For a guessed radial distributions of the density and temperature $(\rho, T)$, 
we solve the radiative transfer equation (Eq. \ref{1218.2}) by the ray-tracing method and obtain 
the Eddington factor $f_{\nu} (r)$ (Eq. \ref{0117.1}). 
For this $(\rho, T)$ distribution, the terms on the right hand side of Equation \eqref{1125.2} except $J_{\nu}$, 
i.e., $B_{\nu}$, $e$, $\Lambda_{\rm chem}$ and  $\Lambda_{\rm line}$, are evaluated. 
Using obtained $f_{\nu} (r)$ and expression for $S_{\nu}$ (Eq. \ref{1125.2} ), 
we calculate the zeroth and the first moments of intensity, $(J_{\nu}, H_{\nu})$, by Equations \eqref{1118.1} and \eqref{1118.2}. 
We then solve Equations \eqref{eq1.6.1}, \eqref{eq1.5} and \eqref{eq1.6} and update the guess for the density and temperature distributions 
using obtained $(J_{\nu}, H_{\nu})$. 
We repeat this procedure until the guessed and obtained $(\rho, T)$ distributions converge anywhere in the envelope.

\section{Escape probability for the H$_2$ line photons}\label{apdi3}

The escape probability $\overline \beta_{\rm esc}$ of H$_2$ line photons 
in Equation \eqref{0113.2} is given by product of that by self-absorption and 
by the continuum absorption: 
\begin{eqnarray}
\overline \beta_{\rm esc} = {\bar f}_{\rm esc} \exp \left( - \tau_{\rm c} \right), \label{ap.2.1}
\end{eqnarray}
where $\tau_{\rm c}$ is the optical depth of the continuum component $\tau_{\rm c}$.
The average escape probability for the lines is calculated for a slab of the H$_2$ column density 
$N_{\rm H_{2}}$ and temperature $T$. 
We consider rovibrational levels in the range $v=0-14$ and $J=0-31$, whose energy 
levels are taken from \citet{Borysow1989} and 
the spontaneous radiative transition coefficients between them are 
from \citet{Wolniewicz1998}.
The level populations are assumed to be in the local thermodynamic equilibrium 
and the photon escape fraction in individual lines is given by 
$f_{{\rm esc}, i}=(1-e^{-\tau_i})/\tau_i$  \citep[see e.g.,][]{Omukai2000}.
The result is fitted in a functional form as
\begin{eqnarray}
{\bar f}_{\rm esc} = \frac{1}{\left(1+N_{\rm H_{2}}/N_{\rm c} \right)^{\alpha}}, \label{ap2.2}
\end{eqnarray}
where 
\begin{eqnarray}
\alpha = a_{0} + a_{1} T + a_{2} T^2 + a_{3} T^3 + a_{4} T^{4} + a_{5} T^5 , \label{ap2.3}
\end{eqnarray}
and 
\begin{eqnarray}
{\rm log} N_{\rm c} = b_{0} + b_{1} T + b_{2} T^2 + b_{3} T^3 + b_{4} T^{4} + b_{5} T^5 . \label{ap2.4}
\end{eqnarray}

The coefficients in equations (\ref{ap2.3}) and (\ref{ap2.4}) are given 
in Tables \ref{a1tab} and \ref{b1tab}.


\bsp	
\label{lastpage}
\end{document}